 \documentclass[final,3p,times]{elsarticle}

\usepackage{graphicx} 
\usepackage{tabularx}
\usepackage{caption}
\usepackage{subcaption}
\usepackage{amssymb}
\usepackage{amsmath}
\usepackage{amsthm}
\usepackage{lineno}
\usepackage[ruled,vlined]{algorithm2e}
\usepackage[export]{adjustbox}
\usepackage{float} 
\usepackage[colorlinks=true, allcolors=blue]{hyperref}
\usepackage[nameinlink]{cleveref}
\usepackage{float}
\usepackage{rotating}
\usepackage{lscape}
\usepackage{fourier} 
\usepackage{array}
\usepackage{makecell}
\usepackage{longtable}
\usepackage{color,soul}
\usepackage{setspace}

\begin{document}

\begin{frontmatter}



\title{Large-Scale Horizontal Axis Wind Turbine Aerodynamic Efficiency Optimization using Active Flow Control and Synthetic Jets}

\author[label1]{Ahmad Nabhani}
\author[label1]{Navid M Tousi}
\author[label2,label3]{Martí Coma}
\author[label4,label5]{Gabriel Bugeda}
\author[label1]{Josep M Bergad\`a\corref{cor1}}
\ead{josep.m.bergada@upc.edu}

\cortext[cor1]{Corresponding author}

\affiliation[label1]{organization={Department of Fluid Mechanics, Universitat Politècnica de Catalunya},
             addressline={Jordi Girona 31},
             postcode={08034},
             state={Barcelona},
             country={Spain}}

 \affiliation[label2]
  {organization={Department of Physics, Universitat Politècnica de Catalunya},
             addressline={Jordi Girona 31},
             postcode={08034},
             state={Barcelona},
             country={Spain}}

\affiliation[label3]
  {organization={Multibody Works S.L.U.},
             addressline={Sant Tomàs 17 B-1},
             postcode={08172},
             state={St Cugat del vallès, Barcelona},
             country={Spain}} 
 
 \affiliation[label4]
  {organization={Department of Civil and Environmental Engineering, Universitat Politècnica de Catalunya},
             addressline={Jordi Girona 31},
             postcode={08034},
             state={Barcelona},
             country={Spain}}

\affiliation[label5]
{organization={International Centre for Numerical Methods in Engineering (CIMNE)},
             addressline={Campus Nord UPC, Edifici C1, Gran Capità},
             postcode={08034},
             state={Barcelona},
             country={Spain}}

\begin{abstract}
Efficiency increase is seen as one of the main goals in any energy converting device. In this direction, the present study aims to demonstrate that large-scale wind turbines can still be improved in order to generate larger amounts of energy. The research presented in this manuscript consists of two main blocks. Initially, it analyzes via Computational Fluid Dynamics (CFD) the boundary layer dynamics on a set of pre-determined airfoils cut along the DTU-10MW reference blade by using the 2D URANS $k-\omega$ SST turbulence model. The aim of this initial stage is to identify the boundary layer separation point, its associated frequency, and peak to peak amplitude for each airfoil cut along the blade, evaluating as well their respective aerodynamic characteristics. The second main goal of the present research consists of implementing the Active Flow Control (AFC) technology and, when employing synthetic jets, to reattach the boundary layer in all airfoils where it is separated. To accomplish this second goal efficiently, the five parameters associated to the AFC implementation performed to each airfoil will be obtained through respective genetic algorithm optimizations. An energy assessment is finally undertaken at each airfoil to validate the respective energy gain obtained. When comparing the net power gain, before and after AFC implementation, generated by each of the airfoils evaluated, net power gains between 23 and 36KW are obtained in all airfoils analyzed, clarifying that the proposed technology is capable of improving the performance of the wind turbines, very likely at any operating condition. 
\end{abstract}


\end{frontmatter}



\section{Introduction}

In recent years, the number of installed large-scale horizontal axis wind turbines  (HAWT) is rapidly growing worldwide. These modern large-scale wind turbines are affected by the unsteady characteristics of the wind turbine operating environment such as turbulence\cite{murata2016experimental, kamada2016effect, ernst2012investigation}, atmosphere stability\cite{sathe2013influence, holtslag2016wind}, gusts\cite{licari_control_2013, li2021gust}, misaligned yaw of the rotor\cite{burton2011wind, del2015estimation}, wind shear\cite{shen2011wind, gould2016effects}, tower shadow\cite{li2018wind, houtzager2013wind}, and upwind wakes\cite{vermeer2003wind, hu2012dynamic, veers2022grand}. All these phenomena will cause the local inflow angle on the rotor blades to change periodically or suddenly\cite{muller2016dynamic, kamada2016effect}. As a result, several airfoils may operate under stall conditions \cite{takao2017experimental, guoqiang2019experimental}, resulting in boundary layer separation, drag increase, loss of lift, deterioration of the airfoils aerodynamic performance \cite{the2017critical, zhu2022combined, zhao2012numerical} and, ultimately, a reduction of wind turbine power output.

Flow control can reduce such unfavorable behavior by delaying the boundary layer separation. Flow control techniques can generally be categorized as passive and active.
Passive flow control techniques do not require input energy and usually utilize geometrical modifications like changing the shape of turbine airfoils via implementing Gurney flaps \cite{zhao2012numerical, storms1994lift, baldacchino2018experimental, wang2008gurney} or vortex generators \cite{zhu2022combined, kerho1993vortex, abbas2013aerodynamic, lin2002review} for example, but their performance is optimal only at design conditions. Wind turbine blades must face a variety of flow conditions during operation, thus it is doubtful whether such control strategies can have a positive effect under all operating conditions. 
In contrast, Active Flow Control (AFC) techniques  (requiring power input) remove the drawback associated with passive control at off-design conditions \cite{jones2018control, rezaeiha2019active,tousi2021active}. This form of control has been associated mainly with adding/subtracting momentum to/from the mean velocity in particular locations in order to interact with the boundary layer and delay/promote its separation. Synthetic Jet Actuators (SJA) \cite{tousi2021active, tousi2022large, amitay2001aerodynamic, buchmann2013influence, zhang2015direct} or direct-barrier-discharge plasma actuators (DBD-PA)\cite{wang2013recent, sato2015mechanisms, de2017micro}, are two of the main devices used in AFC implementations. Further information on flow control actuators can be found in review articles by Cattafesta and Steplak \cite{cattafesta2011actuators}, Johnson et al \cite{johnson2010overview}, and Barlas and Vonkaila \cite{barlas2010review}.

Among the different AFC techniques, synthetic jet actuators have the advantage of generating pulsating flow which combines suction and blowing phases, and they have associated large momentum coefficients. Their high pulsing velocity, fast time response, low power consumption, and easiness of integration to the surface of the body, enable their practical use and maximize the possible aerodynamic gains. Due to the fact that SJA's control is a mature flow control technique of AFC, which is well established and has achieved good results in the field of aeronautics, it is logical to think that their application can be expanded as an effective means to control flow separation on HAWT blades. 
%
%
The application and influence of SJA-AFC on the aerodynamic performance of rotating blades of wind turbines can not be traced until almost the last decade. Stalnov et al. \cite{stalnov2010evaluation} and Maldonado et al. \cite{maldonado2009active,maldonado2010active} were among the first researchers to illustrate experimentally the influence of SJA on the aerodynamic performance of HAWTs.

Stalnov et al. \cite{stalnov2010evaluation} experimentally investigated the performance of SJA based-controlled blades on a modified IAI pr8-SE profile at Reynolds numbers ranging between $2\times 10^5 \leq Re \leq 8\times10^5$ using a Hot-wire sensor. They demonstrated an increase of the overall efficiency of about $5-15\%$. In Maldonado et al. \cite{maldonado2009active, maldonado2010active}, they initially applied SJA to an S809 airfoil in order to control the unsteady loading for a non-rotating and fixed-pitch wind turbine blade. They used 14 SJA in total, distributed in three locations, near the root, around the middle, and at the blade tip. A set of Reynolds numbers ranging between $7.1\times10^4$, and $2.38\times10^5$ were considered, the momentum coefficient range employed was between $C_\mu= 1.34\times10^{-3}$ and $C_\mu= 5.97\times10^{-3}$. 
They demonstrated that the highest influence on flow reattachment and vibration reduction was obtained when placing the SJA near the blade tip. In their second research, the blade was created based on a NACA 4415 airfoil, the number of SJA as well as their location was the same as in their previous research. The Reynolds number range was also the same but the momentum coefficients employed ranged between $9.72\times10^{-4}$ and $1.05\times10^{-2}$. In this second study, they realized the structural vibrations were considerably mitigated thanks to the use of SJA. 

In the research undertaken by Taylor et al. \cite{taylor2015load}, they studied experimentally a WT blade using an S809 airfoil across the entire blade. They incorporated two arrays of 12 SJA placed along the blade and at the normalized chord locations of $x/C=0.1$  and $x/C=0.2$, measured from the leading edge. The Reynolds number was kept constant at $Re=2.2\times10^5$. They demonstrated the airfoil dynamic stall could be displaced to higher Angles of Attack (AoA), and the hysteresis of the lift coefficient when pitching the blade was considerably reduced. A lift to drag ratio hysteresis reduction of around $73\%$ and a significant reduction in flow separation near the leading edge was found in the numerical simulations performed by Tran et al. \cite{tran2014synthetic} on the NREL VI turbine with the S809 airfoil shape at $Re=1\times10^6$ and $C_\mu=0.003$ when the blade was undergoing pitching.  
To investigate the effect of the SJA on the delay of the boundary layer separation and its associated vortical structures, WT studies performed by Rice et al. \cite{rice2019wind} using stereo Particle Image Velocimetry (PIV) and employing the S817 airfoil type were conducted at $Re=3.75\times10^5$. The SJA were placed at $x/C=0.35$, being the momentum coefficients $C\mu=0.012$. When AFC was applied, they observed a significant reduction in the hysteresis of the lift and pitching moment coefficient of $41\%$ and $85\%$, respectively. A numerical study was conducted by Moshfeghi and Hur \cite{moshfeghi2017numerical} on a wind turbine blade 5809 HAWT model for a range of AoA between $0^\circ$ and  $25^\circ$ at $Re=1\times10^6$ under the excitation control of a SJA located at the chord-wise location of $x/C=0.75$. Three different SJA injection angles of $5^\circ$, $15^\circ$ and $25^\circ$ were considered. Their results revealed that for a wide range of injected flow angles, the lift coefficient could be enhanced, no lift coefficient improvement was observed at high AoA and regardless of the SJA injection angle considered.
Maldonado and Gupta \cite{maldonado2019increasing}, experimentally investigated at $Re=1\times10^{6}$ the performance of a real 3-rotor blade based on the S809 airfoil model, they considered five turning speeds, 250, 500, 750, 1000 and 1250rpm, the pitching angles evaluated were $0^{\circ}$, $3^{\circ}$, and $6^{\circ}$. They placed 20 SJA actuators along each blade radius and they used laser Doppler Anomemtry (LDA) to analyze the flow performance at the SJA's exit. For an angular velocity of 500rpm and a pitch angle of $3^{\circ}$, the actuation of 20 SJA per blade decreased the power input to drive the rotor by $10.6\%$, and the rotor efficiency increased by $28\%$. 

Troshin and Seifert \cite{troshin2013performance} studied experimentally at Reynolds numbers ranging from $Re=4*10^5$ to $5\times10^5$, the performance of a thick airfoil AH93-W-300 model typically used at the root section of wind turbines. Three rows of synthetic jets consisting of 13 actuators in each row were considered, the momentum coefficients applied to each SJA were varied from zero (baseline case) to 0.024. A closed-loop control system was used to determine the momentum coefficient applied at each instant. The feedback loop was based on the information provided by the pressure sensors and hot-film sensors placed on the airfoil's upper surface. A significant increase in net energy generated by the wind turbine, up to $60\%$ when compared with the non-actuated performance, was obtained over a wide range AoA and Reynolds numbers. Wu et al. \cite{wu2020role} conducted a numerical study on an elliptic cross-section airfoil at $Re=1100$. They placed a pair of SJA on the upper and lower surfaces and considered the combined effect of pitching and plunging motion. The effect of different parameters such as jet inclination, jet location, and the phase angle between the jet and pitching motion on the energy production capability were analyzed. They observed an increase of up to $30\%$ in energy harvesting along with a significant increase in lift force.
The S809 profile employed in a small HAWT and considering suction to delay the boundary layer separation was studied by \cite{wang2022investigation,sun2024stall}. In both studies, the CFD simulations were performed using the $k-\omega$ SST turbulence model. In the former research the suction groove was located at $0.15C$ versus the leading edge. Four different wind speeds of 7, 10, 13, and $20m/s$ and 12 different momentum coefficients for each wind speed were considered. Thanks to the AFC implementation, the pressure difference between the suction and the pressure surfaces was increased, increasing as well the stability of the flow field. Improvements were particularly relevant at low radius sections, where large boundary layer separation exists. The power capability of the HAWT raised with the wind speed, obtaining an increase versus the base case of $225.56\%$ for a wind speed of $20m/s$. The effectiveness of using a single suction slot or two suction slots combined with a blowing one was analyzed in \cite{sun2024stall}. Slots were located at $0.1C$ and $0.5C$ for the dual case, just the former one was used for the single case. Two momentum coefficients were considered and they observed that when using the larger momentum coefficient and the dual suction configuration, the net power output had a maximum increase. 
Due to its high performance and efficiency, the SD7003 airfoil has been particularly investigated to be employed in small HAWT's and VAWT's, since they operate at relatively low Reynolds numbers $Re \approx 6\times10^4$, \cite{tummala2016review,  muhle2018blind, selig2003low, ducoin2016numerical}. One of the most comprehensive numerical studies to investigate the performance of SJA on this particular airfoil and Reynolds number was undertaken by Tousi et al. \cite{tousi2021active,tousi2022large}. They optimized at four angles of attack (AoA) $4^{\circ}$, $6^{\circ}$, $8^{\circ}$, and $14^{\circ}$ and using an in-house Genetic Algorithms optimizer, the five AFC parameters, jet position, jet with, momentum coefficient, frequency, and jet injection angle. They observed that aerodynamic efficiency could be increased by $251\%$ at $AoA=14^\circ$ while a $39\%$ increase was obtained at $AoA=8^\circ$. In fact, in the present research, the same optimizer and optimization process is employed to obtain the five optimum AFC parameters for different airfoils cut along the DTU-10MW-RWT. 

Based on the existing literature it can be concluded that research on aerodynamic performance of HAWTs with SJA control is rather limited and further knowledge of boundary layer separation/reattachment and its associated dynamics is needed. It is the intention of the authors to bridge this gap by means of the present paper, which is presenting an extensive numerical investigation on 24 baseline case airfoils cut along the blade of a large HAWT. In the second stage, AFC-SJA optimization is implemented to several airfoils. 
In this perspective, the present study clarifies the impact of SJA on the suppression of unsteady separation and dynamic stall on airfoils of HAWT's in order to enhance their aerodynamic performance.

The remainder of the paper is structured as follows. The Governing equations for the URANS turbulence model, along with the computational domain and boundary conditions and the numerical validation are presented in Section 2. The outcome of the baseline simulations studied are introduced in Section 3. The AFC parameters optimization and the results obtained from it are presented in sections 4 and 5 respectively, where the energy assessment is implemented. The paper ends with the main conclusions.  

\section{Numerical method}

\subsection{Governing Equations and Numerical Modeling}

To ensure precise numerical results, selecting an appropriate turbulence model is crucial, especially when dealing with boundary transitions. In this study, we employ the Unsteady Reynolds-Averaged Navier-Stokes (URANS) approach for its computational cost-effectiveness compared to Direct Numerical Simulation (DNS) and Large Eddy Simulation (LES). Specifically, we opt for the $K-\omega$ SST (Shear-Stress Transport) turbulence model \cite{menter1994assessment, menter2003ten}, which excels in handling high Reynolds number scenarios and converges favorably when maintaining a Courant-Friedrichs-Lewy (CFL) number at or below unity $CFL \leq 1$. This choice allows us to investigate critical aerodynamic phenomena, including separation points and stall behavior in FFA-W3-xxx airfoils effectively. We used a finite volume CFD open-source package OpenFOAM to run all the simulations. Also a second order discretization method was used for all parameters for accuracy. To connect pressure and velocity, we chose the Pressure-Implicit with Splitting of Operators (PISO) scheme \cite{issa1986computation, versteeg2007introduction} because it handles dynamic flows well.

The Navier-Stokes equations form the cornerstone of fluid dynamics, describing the conservation of mass and momentum in a continuous fluid medium. In the context of incompressible flow, the only variables associated with the NS equations are the pressure and the three velocity components. Once each variable is substituted by its ensemble-averaged and its fluctuation terms, and after using the Boussinesq approximation, the Navier-Stokes equations can be expressed as:

\begin{equation}
\nabla \cdot \mathbf{\overline{u}} = 0
\end{equation}


\begin{equation}
\rho \frac{\partial \mathbf{\overline{u}}}{\partial t} + \rho \mathbf{\overline{u}} \cdot \nabla \mathbf{\overline{u}} = -\nabla \overline{p} + (\mu + \mu_t) \nabla^2 \mathbf{\overline{u}} 
\end{equation}

Where, $\mathbf{\overline{u}}$ and $\overline{p}$ respectively represent the ensemble-averaged velocity vector and pressure, $\mu$ is the kinematic viscosity and $\mu_t$ characterizes the apparent or turbulence viscosity.

In our current investigation, we opted to employ the $k-\omega$ SST turbulence model \cite{menter1994assessment, menter2003ten}. This model incorporates the $k-\omega$ formulation in proximity to the wall, the $k-\epsilon$ model in regions far from the object, and a blending function bridging these two approaches. Expressing the turbulence viscosity $\mu_t$ mathematically for the chosen model yields:
\begin{equation*}
    \mu_t = \frac{\rho k}{\omega} \quad \longrightarrow \quad 
    \begin{cases}
        \rho: \text{density} \\
        k: \text{turbulent kinetic energy} \\
        \omega: \text{turbulent kinetic energy specific dissipation rate} 
    \end{cases}
\end{equation*}

This model relies on two transport equations to solve for $k$ and $\omega$. The governing equations for each parameter are given by:
\begin{equation}
    \frac{\partial k}{\partial t}+u_j\frac{\partial k}{\partial x_j}=P_k-\beta^*k\omega +\frac{\partial }{\partial x_j}\left [ \left ( \nu +\sigma_k\nu_T \right ) \frac{\partial k}{\partial x_j} \right ]
\end{equation}

\begin{equation}
    \frac{\partial \omega}{\partial t}+u_j\frac{\partial \omega}{\partial x_j}=\alpha S-\beta\omega^2 +\frac{\partial }{\partial x_j}\left [ \left ( \nu +\sigma_\omega \nu_T \right ) \frac{\partial \omega}{\partial x_j} \right ]+2\left ( 1-F_1 \right )\sigma_{\omega 2}\frac{1}{\omega}\frac{\partial k}{\partial x_i}\frac{\partial \omega}{\partial x_i}
\end{equation}

The blending functions $F_1$ and $F_2$ are contingent on the distance from the cell to the wall. $F_1$ equals 0 far from the wall, 1 near the wall, and varies between 0 and 1 in the transition region. Meanwhile, $F_2$ relies on the perpendicular distance from the wall ($d$), expressed mathematically as:
\begin{equation}
    F_2 = \tanh \left ( \arg^2_2 \right )
\end{equation}
\begin{equation}
    \arg^2_2 = \max\left ( \frac{2k}{\beta^*\omega d};\: \frac{500 \nu}{\omega d^2} \right )
\end{equation}

The constants of the $\arg$ function are manually adjusted, and in our specific case, the turbulence model employs the following values: $\alpha_1=0.556$, $\alpha_2=0.44$, $\beta^*=0.09$, $\beta_1=0.075$, $\beta_2=0.0828$, $\sigma_{k1}=0.85$, $\sigma_{k2}=1$, $\sigma_{\omega1}=0.5$, $\sigma_{\omega2}=0.856$. For an in-depth understanding of the proposed $k-\omega\:SST$ turbulence model, refer to \cite{menter1994assessment}.

\subsection{Wind Turbine Design and Key Parameters}

\begin{figure}[t]
    \centering
    \begin{subfigure}[b]{0.49\textwidth}
        \centering
        \includegraphics[width=\textwidth]{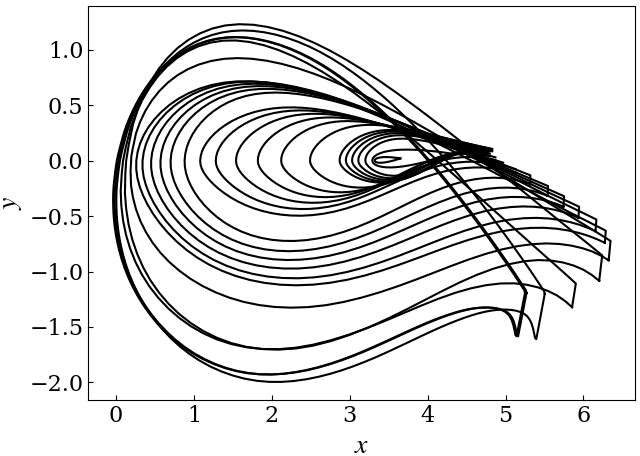}
        \caption{}
        \label{fig-d-a}
    \end{subfigure}
    \begin{subfigure}[b]{0.41\textwidth}
        \centering
        \begin{subfigure}[b]{\textwidth}
            \centering
            \includegraphics[width=\textwidth]{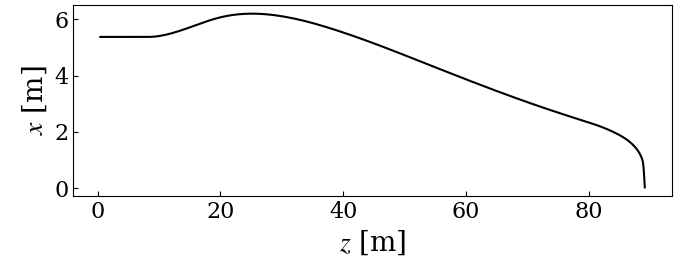}
            \caption{}
            \label{fig-d-b}
        \end{subfigure}
        \begin{subfigure}[b]{\textwidth}
            \centering
            \includegraphics[width=\textwidth]{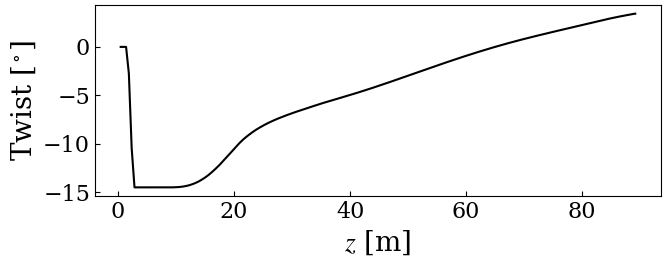}
            \caption{}
            \label{fig-d-c}
        \end{subfigure}
    \end{subfigure}
    \caption{Key design dimensions. Airfoil sections considered (a), chord variation along the blade radius (b) and twist angle associated to each section (c).}
    \label{fig-design}
\end{figure}

The DTU-10MW, a reference wind turbine developed by Denmark Technical University, is employed in the current study due to its open-source nature and its inherent suitability for researchers in the field. The DTU-10MW turbine represents a noteworthy embodiment, characterized by a three-bladed upwind horizontal-axis configuration, variable-speed operation, and pitch-regulated yaw control. The comprehensive geometrical and operational attributes of this wind turbine, encompassing critical aerodynamic, structural, and control specifications, are meticulously documented in \Cref{table01} and \Cref{table_02}. In the first table, the main dimensional characteristics of the WT are introduced. \Cref{table_02} presents the six airfoil types employed in the construction of the blade. The maximum non dimensional thickness of each airfoil $(t/C)_{max}$ and the non dimensional radial location $z/R$ where each airfoil is employed are respectively shown in the second and third columns. The final column characterizes the maximum and minimum angles of attack employed in each section.
To complement the information presented in these tables, \Cref{fig-design}a introduces the profile of the 24 sections evaluated in the present study. \Cref{fig-design}b and \Cref{fig-design}c, respectively show the airfoil chord variation along the blade radius and the twist angle associated with each airfoil. 

Furthermore, this study focuses on two pivotal parameters that underpin the requisite configuration for our simulation endeavors: the Reynolds number distribution along the blade and the chord length distribution along the blade. These parameters, play an indispensable role in facilitating the precise modeling and analysis of the turbine's dynamic behavior under varying operational scenarios. For a more exhaustive exploration of the DTU-10 MW wind turbine's intricacies and design particulars, interested readers are encouraged to consult the comprehensive exposition provided in Reference \cite{bak2013description}. It is imperative to emphasize that this turbine's geometric and operational facets, spanning aerodynamic profiles, structural attributes, and control systems, serve as the foundational inputs that propel our simulation analyses and underpin the academic pursuits undertaken in this study.

\begin{table}[t]
\centering
\caption{Key parameters of the DTU 10 MW Reference Wind Turbine. \cite{bak2013description}}
\begin{tabular}{l*{2}{c}r} 
\hline
Parameter &    \\
\hline
Rotor Orientation &  Clockwise rotation \\
Cut in wind speed & 4 [m/s]  \\
Cut out wind speed & 25 [m/s]  \\
Rated wind speed &  11.4 [m/s]  \\
Number of blades &  3   \\
Blade radius &  89.17 [m] \\
Hub height &  119 [m] \\
\hline
\label{table01}
\end{tabular}
\end{table}

\begin{table}[b]
\centering
\caption{DTU-10MW airfoils and design values \cite{bak2013description}.  }
\begin{tabular}{{p{3cm} p{1.5cm} p{2cm} p{3cm} p{2cm}}} 
\hline
 Airfoil type  & $(t/C)_{max}$ &  z/R & z(m) &   $\alpha(^\circ)$  \\
\hline
FFA-W3-600 & 0.6 & 0.04-0.16 & $4.4\leq r \leq 14.2$ & $62^\circ-36^\circ$\\
FFA-W3-480 & 0.48 &  0.16-0.25 & $14.2 \leq r \leq 22.2$ & $35^\circ-23^\circ$\\
FFA-W3-360 & 0.36 &  0.25-0.34 & $22.2 \leq  r \leq 30.2$ & $22^\circ-18^\circ$\\
FFA-W3-301 & 0.301 &  0.34-0.6 & $30.2 \leq r \leq 53.4$ & $17^\circ-10^\circ$\\
FFA-W3-241 & 0.241 &  0.6-0.99 & $53.4 \leq r \leq 89$ & $9^\circ-0.9^\circ$\\
NACA0015 & 0.15 &  0.99-1 & $89 \leq r \leq 89.17$ & $0^\circ$\\
\hline
\label{table_02}
\end{tabular}
\end{table}

\subsection{Domain and Boundary Conditions}

\begin{figure}[t]
     \centering
     \begin{subfigure}[b]{0.5\textwidth}
         \centering
         \includegraphics[width=\textwidth]{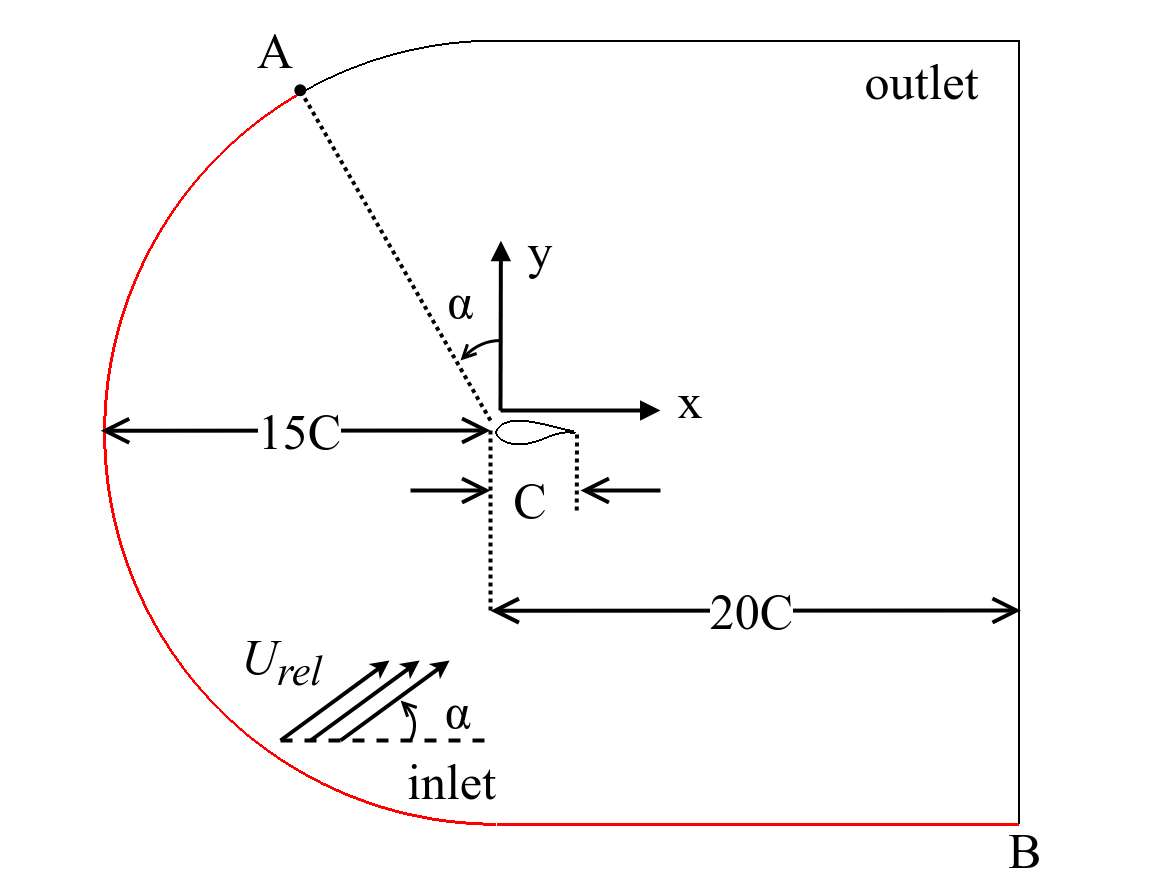}
         \caption{}
         \label{fig1a}
     \end{subfigure}
     \begin{subfigure}[b]{0.45\textwidth}
         \centering
         \includegraphics[width=\textwidth]{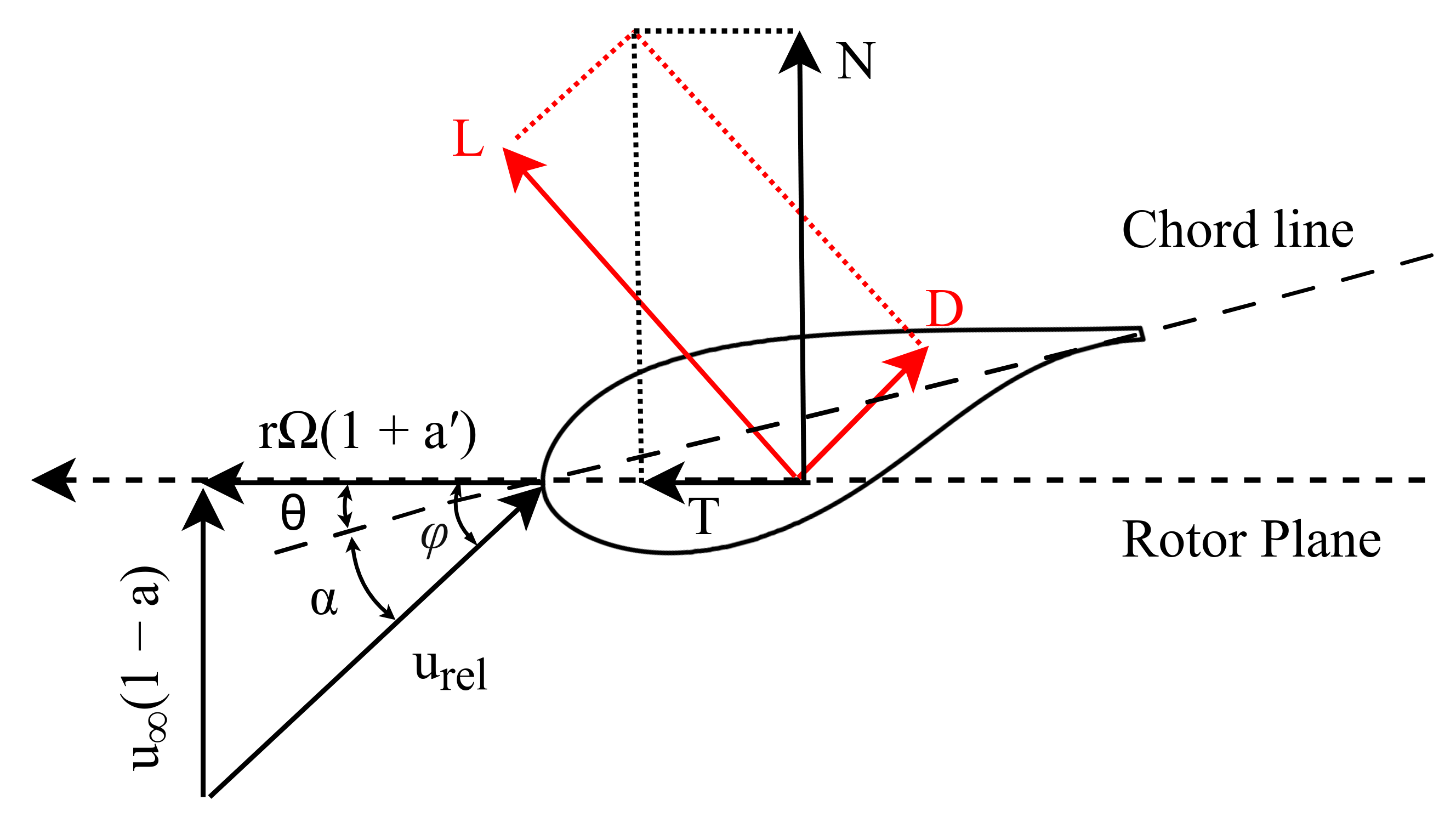}
         \caption{}
         \label{dig1b}
     \end{subfigure}
     \hfill
     \caption{2D Computational domain (a). Angles and forces associated to a given airfoil belonging to a HAWT (b).}
\label{fig-domain}
\end{figure}

\Cref{fig1a}, introduces the computational domain used to perform all CFD simulations presented in this document. The origin of the coordinate system was placed at the corresponding airfoil leading edge. The computational domain upstream and downstream boundaries were set to 15 and 19 times the chord length (C), respectively. Each airfoil section was placed horizontally in the computational domain, the Angle of Attack (AoA) was set to zero degrees. To be able to consider the required AoA for each airfoil, the corresponding components of the relative velocity \( (u, v) = (U_{\text{rel}} \cos \alpha, U_{\text{rel}} \sin \alpha) \) were set at the computational domain inlet. 


Where \( U_{\text{rel}} \) is the relative speed, calculated using the blade element momentum theory \cite{rankine1865mechanical, froude1889part, wilson2018applied,hansen2015aerodynamics, burton2011wind}.
\begin{equation}
U_{\text{rel}} = \sqrt{\left( r  
\Omega  (1 + a') \right)^2 + \left(U_\infty (1 - a) \right)^2}
\end{equation}

The angular velocity of the rotor (see figure \ref{dig1b}) is defined as (\( \Omega \)), the effective angular induction factor is given as (\( a' \)), the free stream wind velocity is represented as (\( U_\infty \)), the parameter (\( a \)) characterizes the axial induction factor. The local angle of attack (AoA) is represented by $\alpha=\varphi-\beta$, where $\varphi$ or inflow angle is the angle between the plane of rotation and the relative speed, $U_{rel}$, and $\beta$ is the local twist angle of the blade. A possible relation between these parameters is: 
\begin{equation}
    tan\varphi=\frac{(1-a)U_\infty}{(1+a') r  \Omega}
\end{equation}

It is recalled from each section considering the 2D aerodynamics that the lift, by definition, is perpendicular to $U_{rel}$ while the drag is parallel to the same velocity. Furthermore, the lift $L$ and drag $D$ forces associated to each airfoil can be found in equations (9) and (10):

\begin{equation}
    L=\frac{1}{2} \rho U^2_{rel} C C_l
\end{equation}

\begin{equation}
    D=\frac{1}{2} \rho U^2_{rel} C C_d 
\end{equation}
C being the airfoil chord, $\rho$ the fluid density and $C_L$ and $C_D$ the lift and drag coefficients respectively.
Since we are interested in the forces tangential $F_{Tr}$ and normal $F_N$ to the rotor plane, the Lift, and Drag forces have to be projected versus these two respective directions:

\begin{equation}
    F_{Tr} = LSin\varphi - DCos\varphi
\end{equation}

\begin{equation}
    F_N = LCos\varphi + DSin\varphi
\end{equation}

The boundary conditions for the velocity were Dirichlet at the computational domain inlet and at the synthetic jet groove, for the cases where AFC was considered. Velocity was set as null at the airfoil surface and Newmann boundary conditions were set at the computational domain outlet. 
The time-dependent, top hat, velocity profile employed at the synthetic jet actuator was defined as: 
\begin{equation}
   u_j = U_j \sin(2\pi F t) 
\end{equation}
where $U_j$ is the maximum jet velocity, $F$ is the jet frequency, and $t$ is the dimensional time.

Neumann boundary conditions for pressure were used at the computational domain inlet, synthetic jet, and airfoil walls, and Dirichlet boundary conditions for pressure were used at the computational domain outlet. Note that the outlet is represented by the upper curve from points $A$ to $B$ in \Cref{fig1a}, wound clockwise. The computational domain inlet is represented in the same figure from points $B$ to $A$, in a clockwise direction.

The values of the turbulence parameters associated with the turbulence model chosen for this paper, needed to be determined at the computational domain inlet based on the following equations: $k=(3/2)U_{rel}^2 
I^2$; $\epsilon=C_{\mu}(k^{3/2}/C)$;  where $C_{\mu}=0.09$; 
$\omega=\epsilon/(C_{\mu}k)$; and $\nu_t=k/\omega$. Using these equations, values 
for the turbulent 
kinetic energy $k$, the turbulent dissipation rate $\epsilon$, the specific dissipation range 
$\omega$, and the turbulent kinematic viscosity $\nu_t$ at the computational domain inlet were determined for each case. The turbulence intensity $I$ at the computational domain inlet was kept constant and equal to $I=0.1\%$ for all cases studied. Newmann boundary conditions for all turbulence parameters were set at the computational domain outlet and at the synthetic jet. A Dirichlet value of $10^{-20}$ for the turbulence kinetic energy was employed at the airfoil walls.  

\subsection{Non dimensional groups}

The lift coefficient ($C_l$) is a measure of the lift force generated by any fluff-body in the direction perpendicular to the flow velocity. 
\begin{equation}
C_l = \frac{L}{0.5 \rho U_{rel}^2 A}
\end{equation}
where $L$ is the lift force, $\rho$ is the fluid density, $U_{rel}$ is the flow relative velocity, and $A$ is the reference area (typically the airfoil chord for 2D simulations).

The drag coefficient ($C_d$) quantifies the drag or resistance of an object in a fluid environment.
\begin{equation}
C_d = \frac{D}{0.5 \rho U_{rel}^2 A}
\end{equation}
$D$ being the drag force.

The aerodynamic efficiency is the ratio between the lift and drag coefficients:
\begin{equation}
    \eta = \frac{C_l}{C_d}
    \label{eq:E}
\end{equation}

The pressure coefficient ($C_p$) characterizes the strength of the pressure field around any bluff-body and it is defined as:
\begin{equation}
C_p = \frac{p - p_\infty}{0.5 \rho U_{rel}^2}
\end{equation}
where $p$ is the local pressure and $p_\infty$ is the free-stream pressure.

The skin friction coefficient ($C_f$) measures the skin friction produced by viscous forces at the surface of a body immersed in a fluid.
\begin{equation}
C_f = \frac{\tau_w}{0.5 \rho U_{rel}^2}
\end{equation}
where $\tau_w$ defines the wall shear stresses.

The Reynolds number is defined as the ratio of inertial to viscous forces. Mathematically it is expressed as:
\begin{equation}
    Re=\frac{\rho U_{rel} C}{\mu}
    \label{eq:Reynolds}
\end{equation}
where $\mu$ defines the fluid absolute viscosity.

The momentum coefficient represents the momentum associated with the jet divided by the incoming fluid momentum. According to reference, \cite{goodfellow2013momentum} $C_{\mu}$ is the primary AFC parameter. Mathematically it is expressed as:
\begin{equation}
    C_\mu = \frac{h\left ( \rho U_{j}^2  \right )\sin \theta_{\text{jet}}}{C\left ( \rho U_{rel}^2  \right )}
    \label{eq:Cmu}
\end{equation}
where ($h$) is the jet width, ($U_{j}$) is the maximum jet velocity and ($\theta_{\text{jet}}$) is the jet inclination angle with respect to the airfoil surface.

The forcing frequency given as a non dimensional number is defined as follows:
\begin{equation}
    F^+= \frac{f}{f_0}
    \label{eq:forcfreq}
\end{equation}
where $f$ is an arbitrary frequency and $f_0$ is the vortex shedding frequency (the frequency at which vortices are shed from the airfoil).

Y Plus ($y^+$) defines the non-dimensional distance from the wall to the first grid cell, normalized by the viscous length scale.
\begin{equation}
y^+ = \frac{y u_\tau}{\nu}
\end{equation}
where $y$ is the physical distance from the wall to the cell, $u_\tau$ is the friction velocity ($u_\tau = \sqrt{\frac{\tau_w}{\rho}}$), and $\nu$ is the kinematic viscosity.

The Courant-Friedrichs-Levy number, often denoted as $\mathrm{CFL}$ refers to the ability of the system to capture the field of interest. Mathematically it is defined as the fluid velocity measured at any given mesh cell multiplied by the simulation time step and divided by the mesh cell length scale ($\Delta x$). 
\begin{equation}
    CFL = \frac{{u} \Delta t}{\Delta x}
    \label{eq:CFL}
\end{equation}
For time-dependent simulations, the Courant number is an important parameter in order to ensure stability of the system, and it is recommended to keep it below 1.

\subsection{Mesh Sensitivity Study and Numerical Validation}

\begin{figure}[t]
     \centering
     \begin{subfigure}[b]{0.30\textwidth}
         \centering
         \includegraphics[width=\textwidth]{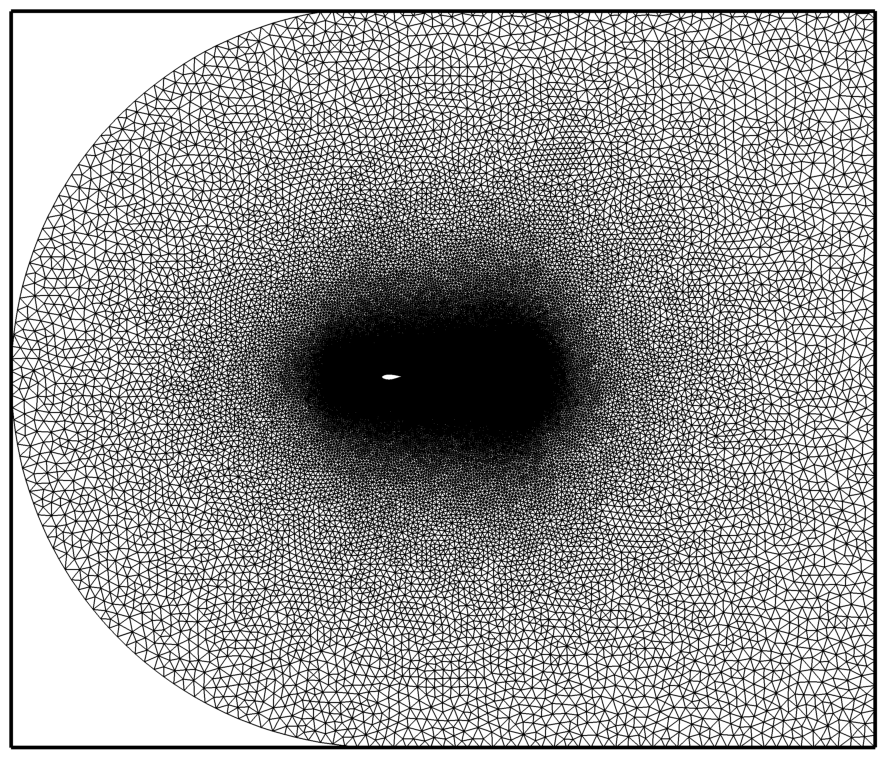}
         \caption{ }
         \label{fig01-a}
     \end{subfigure}
     \hfill
     \begin{subfigure}[b]{0.30\textwidth}
         \centering
         \includegraphics[width=\textwidth]{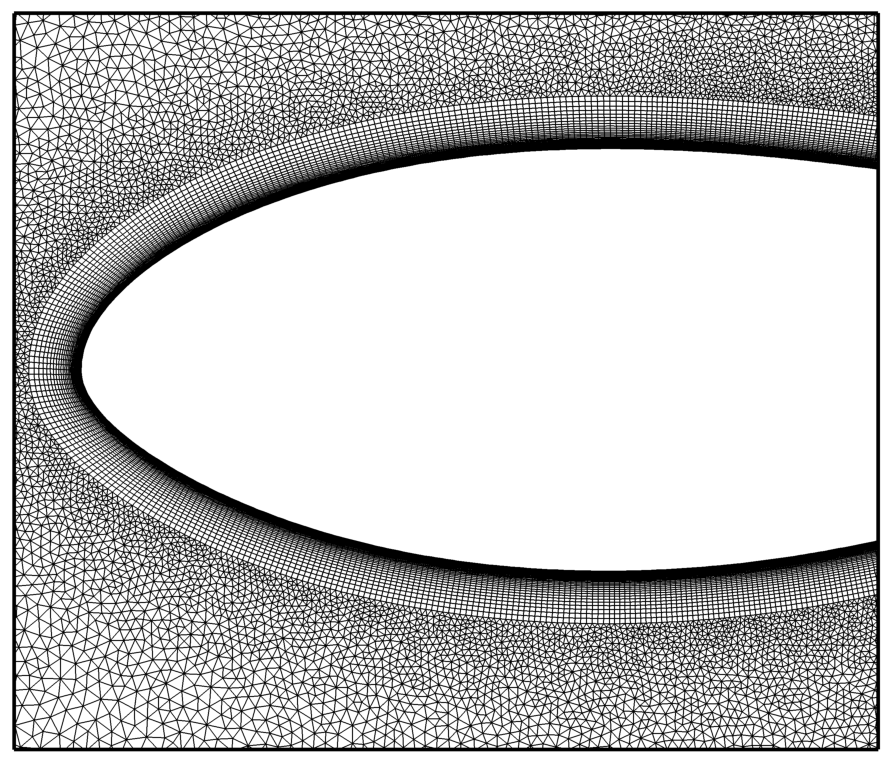}
         \caption{ }
         \label{fig01-b}
     \end{subfigure}
     \hfill
     \begin{subfigure}[b]{0.30\textwidth}
         \centering
         \includegraphics[width=\textwidth]{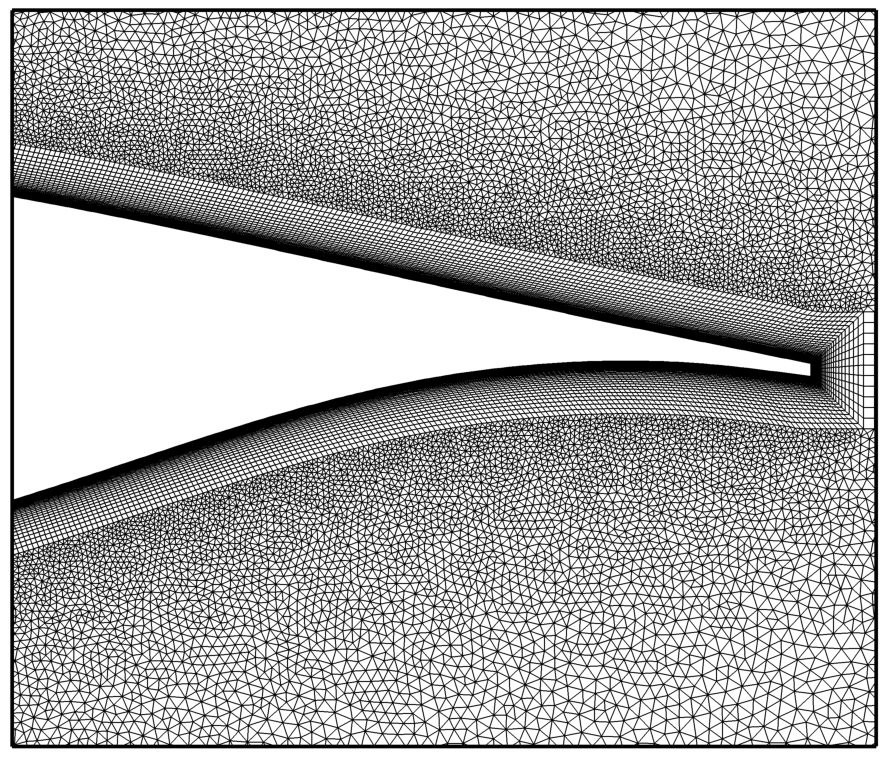}
         \caption{ }
         \label{fig01-c}
     \end{subfigure}
     \hfill
     \caption{(a) General view of the complete mesh, (b) leading edge and (c) trailing edge zoomed views of the mesh.}
\label{figure01}
\end{figure}

In order to make sure the results are fully accurate, we meticulously conducted a mesh sensitivity analysis using the airfoil located at $z/R = 0.45$, where for a wind speed of $10 m/s$ the Reynolds number is the highest along the blade ($1.53*10^{7}$). In this particular section, the airfoil type is the $FFA-W3-301$, being the associated relative speed and the angle of attack (AoA) of $41.2 m/s$ and $14.22^\circ$, respectively. 
To be able to capture the boundary layer dynamics and the time averaged separation point while maintaining a reasonable computational time we decided to use a hybrid mesh. This allowed us to refine the mesh near the airfoil without significantly increasing the overall number of cells. An overview of the mesh in the entire computational domain is presented in \Cref{fig01-a}, while \Cref{fig01-b,fig01-c} provide close-up views of the mesh refinement near the airfoil's leading and trailing edges, respectively.

\begin{figure}[t]
    \centering
    \begin{subfigure}[b]{0.45\linewidth}
    \includegraphics[width=\linewidth]{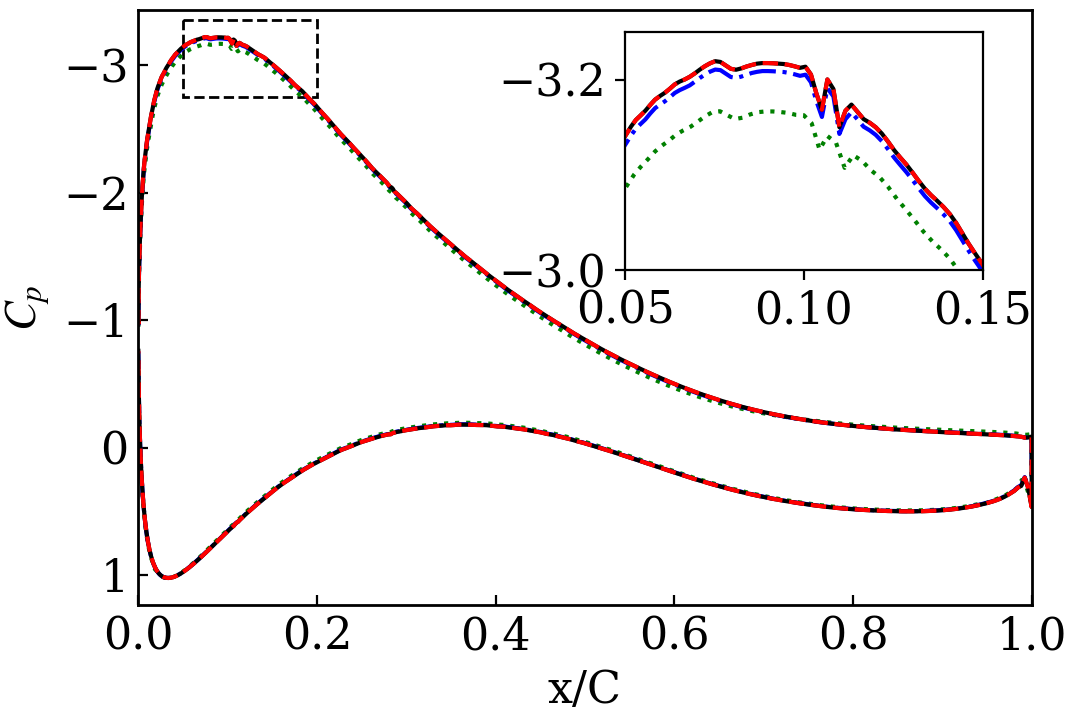}
    \caption{ }
    \end{subfigure}
    \begin{subfigure}[b]{0.47\linewidth}
    \includegraphics[width=\linewidth]{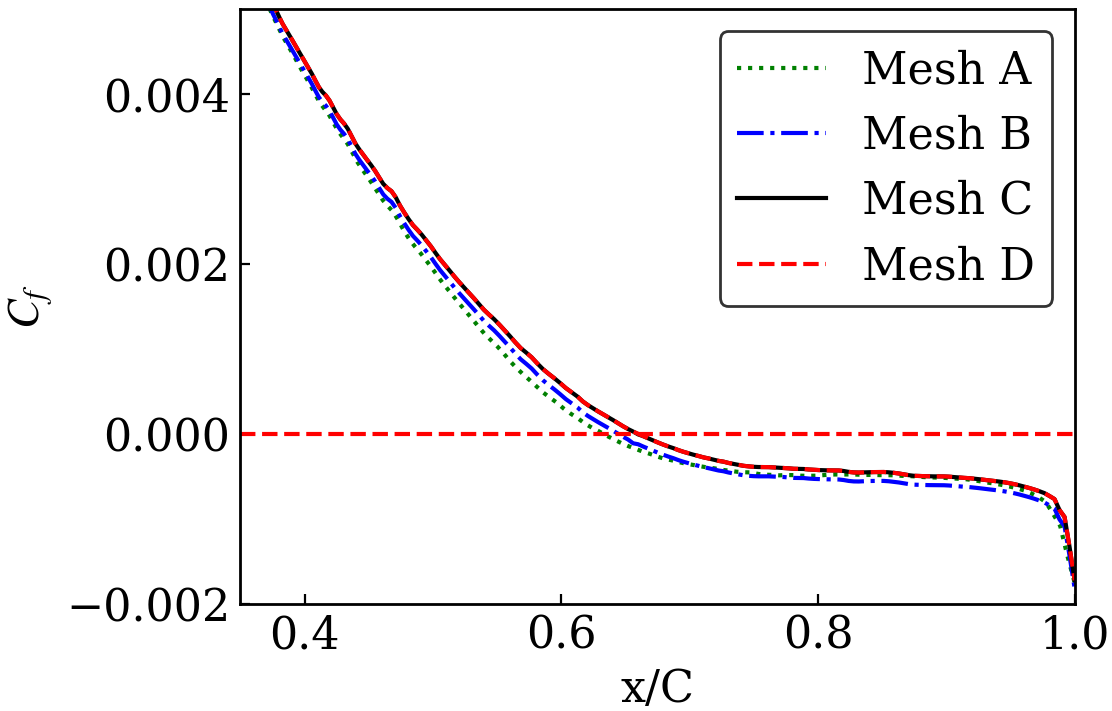}
    \caption{ }
    \end{subfigure}
\caption{distribution of (a) pressure coefficient (for suction and pressure surfaces) with a zoomed view and (b) Skin friction coefficient for (suction surface), for $z/R=0.45$.}
\label{figure04}
\end{figure}

We examined four different meshes labeled A, B, C, and D, which details are introduced in \Cref{table02}. The coarsest mesh had $200K$ cells, whereas the finest had $323K$ cells. The third column of the table displays the estimated $y+$ value for the first cell layer from the airfoil surface, the fourth column is presenting the maximum $y+$ value after the simulations. The subsequent column reveals the location along the chord where separation occurs $(x_s/C)$. This separation point is crucial for assessing the AFC implementation. Then it has a significant impact on the optimum AFC parameters. 
Columns 6 through 10 of \Cref{table02} report the lift and drag coefficients, their respective errors versus the finest mesh values, and the aerodynamic efficiency of the airfoil. The last column displays the vortex shedding frequency which is obtained from the FFT of the dynamic lift coefficient. With the exception of mesh A, all meshes yield similar aerodynamic performance characteristics. Additionally, the pressure $C_p$ and friction $C_f$ distributions, see \Cref{figure04}, are practically identical among meshes B, C, and D. Quantitatively, mesh C achieves three significant digits accuracy in all considered parameters when compared versus the maximum resolution values of mesh D. This makes mesh C an excellent compromise between accuracy and computational requirements, consisting of around 300K hybrid cells in total and 43K quadrilateral cells in the halo region, which has 52 layers in the direction perpendicular to the surface and 840 cells along the airfoil surface. The first layer thickness maintains a maximum $y^+$ below $0.71$, meeting the requirements of the $K-\omega$ SST turbulence model wall treatment to calculate shear stress and find the correct boundary layer separation point.

\begin{table}[t]
\caption{Time-averaged aerodynamic coefficients for different meshes, wind speed of $10[m/s]$ and $z/R=0.45$.}
\begin{center}
\begin{tabular}{p{0.8cm} p{1.1cm} p{0.6cm} p{1.1cm} p{1.1cm} p{1.1cm} p{1.2cm} p{1.1cm} p{1.1cm} p{1.9cm} p{1.5cm} } 
\hline
Mesh & $N_{cell}$ & $y^+$ & $Cal. y+$ & $x_s/C$ & $C_l$ & $\epsilon_{C_l}$ [\%] & $C_d$  & $\epsilon_{C_d} [\%]$ & $\eta=C_l/C_d $ & F. [Hz]  \\
\hline
A &  $200156$ & $4$ & $3.79$ & $0.63538$ & $1.2812$& 1.31 & $0.04975$ & 2.40 &$25.75$ & $9.3$  \\
B & $271632$ &$1$ &  $0.98$ & $0.64817$ & $1.2908$& 0.7 & $0.04881$ & 0.47 & $26.44$ & $8.5$  \\
C & $300698$ &$0.7$ &  $0.71$ & $0.6599$ &  $1.3023$& - & $0.04862$ & 0.08 & $26.78$ & $7.52$  \\
D &  $323009$ &$0.5$ & $0.49$ & $0.6599$ &  $1.3023$& - & $0.04858$ & - & $26.80$ & $7.52$  \\
\hline
\end{tabular}
\end{center}
\label{table02}
\end{table}

We tested three time steps $1\times10^{-5} (s)$, $5\times10^{-6} (s)$, and $2\times10^{-6} (s)$, obtaining identical results for converged solutions. For the baseline cases, we decided to use a time step of $5\times10^{-6} (s)$, but to resolve AFC cases accurately, we selected the lowest time step $2\times10^{-6} (s)$, in order to keep the CFL number below one. The convergence criterion was set to $10^{-6}$ for all parameters and in all simulations. Convergence was consistently achieved before 10 time units, and the maximum time horizon for all simulations was set to $T=30(s)$ to ensure a robust field average.

\Cref{table03} presents a further comparison between the present work and the research undertaken by \cite{zahle2014comprehensive}, in which two different 3D-CFD methodologies were employed to simulate the full 3D blade of the DTU-10MW HAWT. The comparison between the previous and the present research was performed on the airfoil located at $z/R=0.45$, since it is the one considered in our mesh independence study. In this table, the time averaged lift and drag coefficients as well as the airfoil aerodynamic efficiency are presented. The error between the present results and the ones gathered by \cite{zahle2014comprehensive} show a good agreement, further validating the present 2D simulations. 

\begin{table}[b]
\caption{Aerodynamic coefficients of the Baseline case airfoil at z/R=0.45, comparison with the work done by  \cite{zahle2014comprehensive}.}
\centering
\begin{tabular}{llllllll}
\hline
$z/R$ & Cases & $C_l$ & $\epsilon_{C_l}$ & $C_d$ & $\epsilon_{C_d}$ & $\eta$  & $\epsilon_{\eta}$ \\
\hline
0.45    &  BEM CFD 3D \cite{zahle2014comprehensive} & 1.2016  &  - &  0.04301 & - & 27.93  & -  \\
    & Present $K-\omega$ SST (2D) & 1.3023 &  8.38 \%  &  0.04862 &  1.30 \% & 26.78  & 4.11 \% \\
    \cline{2-8}
    & ElipSys 3D \cite{zahle2014comprehensive} & 1.2616 & - & 0.04809 & - &  26.23 & - \\
    & Present $K-\omega$ SST (2D) & 1.3023 &  3.22 \%  &  0.04862 &  1.08 \% &  26.78 & 2.09 \% \\
\hline
\end{tabular}
\label{table03}
\end{table}

\section{Baseline computations and results}

There is very little information based on experimental testing or CFD simulations for large Horizontal Axis Wind Turbines (HAWTs). In fact, there is one previous 3D-CFD study on the DTU-10MW wind turbine operating at a wind speed of $10m/s$ done by Zahle \cite{zahle2014comprehensive}, where he used the BEM and EllypSys3D methods. In \Cref{table03} we used Zahle's work as a reference to make sure our boundary conditions and mesh are correct. In the present section the comparison will be extended to the rest of the sections studied.

\begin{figure}[t]
     \centering
     \begin{subfigure}[b]{0.48\textwidth}
         \centering
        \includegraphics[width=\textwidth]{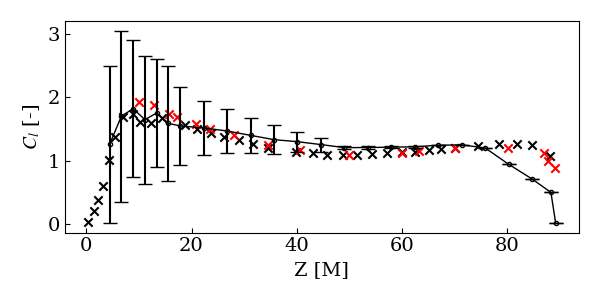}
         \caption{}
        \label{fig5-a1}
     \end{subfigure}
          \begin{subfigure}[b]{0.48\textwidth}
         \centering
         \includegraphics[width=\textwidth]{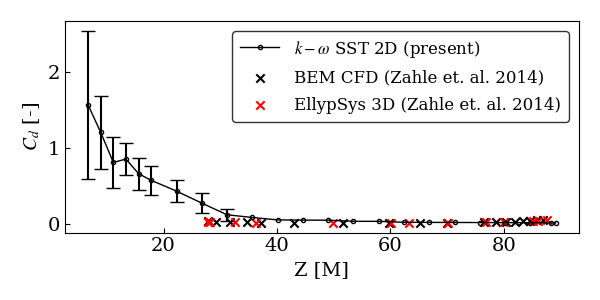}
         \caption{}
        \label{fig5-b1}
     \end{subfigure}
     \caption{Comparison of (a) Lift coefficient distribution (b) Lift coefficient distribution along the blade(Error bars showing the amplitude of vortex shading) with previous works \cite{zahle2014comprehensive}.}
        \label{fig5_1}
\end{figure}

\Cref{tab_all_sections} in Appendix, summarizes the main information of the 24 sections (airfoils) we have divided the WT blade under study. Each airfoil has been carefully studied via 2D-CFD and the airfoil type and location, the AoA associated, the chord, the time-averaged lift and drag coefficients and the boundary layer separation point $x_S/C$, are presented in the table. In fact, at the same time averaged $C_l$ and $C_d$ values are as well presented in \Cref{fig5_1}, where these coefficients at the different airfoils are compared with the ones obtained by \cite{zahle2014comprehensive}. The agreement is very good although some discrepancies are observed nearby the blade tip, probably due to the particularly relevant three-dimensional flow effects which may not have been properly gathered in our 2D-CFD simulations. In the same figure, we have implemented the $C_l$ and $C_d$ peak-to-peak amplitude associated with each airfoil, amplitudes associated with the vortex shedding, and the boundary layer dynamic displacement. Note that these amplitudes are particularly large nearby the root where an early boundary layer separation is expected. Amplitudes keep decreasing as we move towards the blade tip.
By examining the lift and drag coefficients in all airfoils, we can determine the torque and thrust force generated by the turbine. These forces, in turn, contribute to the overall power output of the turbine. Essentially, the lift and drag coefficients provide valuable insights into the aerodynamic performance of each turbine airfoil, aiding in assessing its efficiency and power generation capabilities.

\begin{figure}[t]
    \centering
    \begin{subfigure}[b]{0.7\textwidth}
    \includegraphics[width=\textwidth]{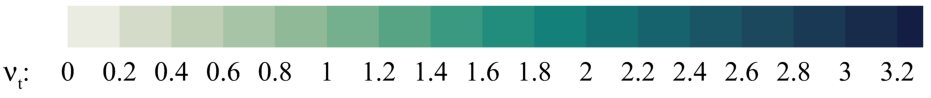}
     \end{subfigure}
    \begin{subfigure}[b]{0.3\textwidth}
    \includegraphics[width=\textwidth]{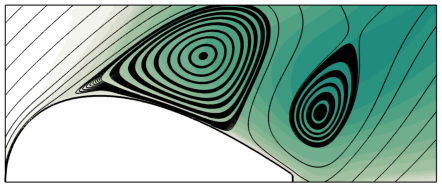}
    \caption{z/R = 0.10}
     \end{subfigure}
    \begin{subfigure}[b]{0.3\textwidth}
    \includegraphics[width=\textwidth]{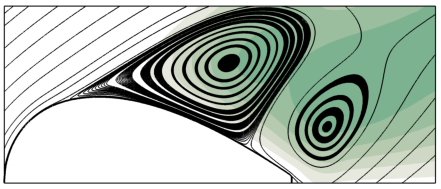}
    \caption{z/R = 0.15}
     \end{subfigure}
     \begin{subfigure}[b]{0.3\textwidth}
     \includegraphics[width=\textwidth]{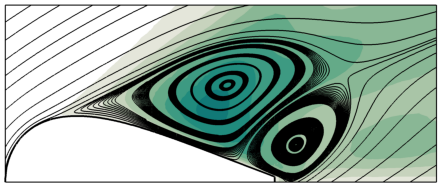}
    \caption{z/R = 0.20}
     \end{subfigure}
    \begin{subfigure}[b]{0.3\textwidth}
    \includegraphics[width=\textwidth]{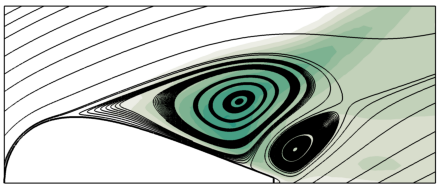}
    \caption{z/R = 0.25}
     \end{subfigure}
     \begin{subfigure}[b]{0.3\textwidth}
     \includegraphics[width=\textwidth]{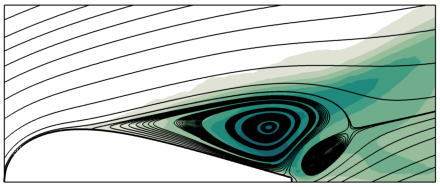}
    \caption{0.30}
     \end{subfigure}
    \begin{subfigure}[b]{0.3\textwidth}
    \includegraphics[width=\textwidth]{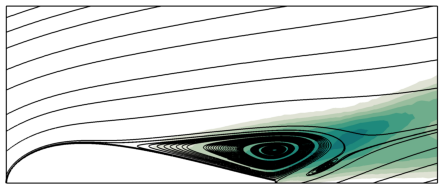}
    \caption{z/R = 0.35}
     \end{subfigure}
     \begin{subfigure}[b]{0.3\textwidth}
     \includegraphics[width=\textwidth]{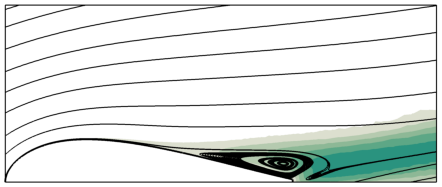}
    \caption{z/R = 0.40}
     \end{subfigure}
     \begin{subfigure}[b]{0.3\textwidth}
     \includegraphics[width=\textwidth]{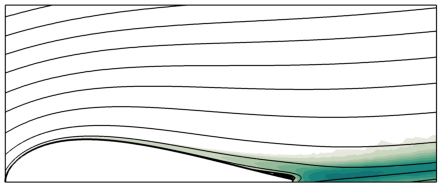}
    \caption{z/R = 0.50}
     \end{subfigure}
     \begin{subfigure}[b]{0.3\textwidth}
     \includegraphics[width=\textwidth]{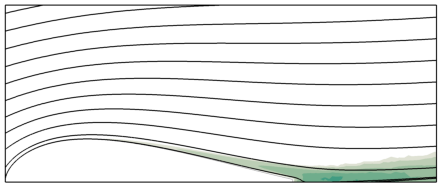}
    \caption{z/R = 0.60}
     \end{subfigure}
     \begin{subfigure}[b]{0.3\textwidth}
     \includegraphics[width=\textwidth]{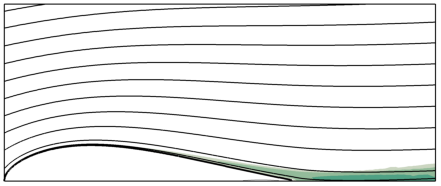}
    \caption{z/R = 0.70}
     \end{subfigure}
     \begin{subfigure}[b]{0.3\textwidth}
     \includegraphics[width=\textwidth]{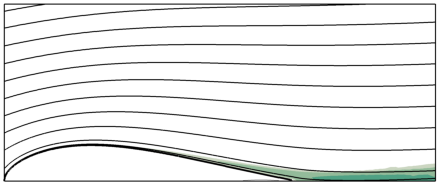}
    \caption{z/R = 0.80}
     \end{subfigure}
     \begin{subfigure}[b]{0.3\textwidth}
     \includegraphics[width=\textwidth]{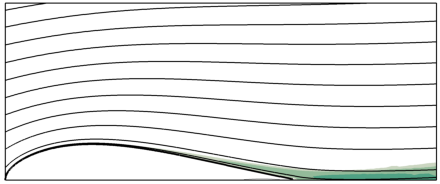}
    \caption{z/R = 0.90}
     \end{subfigure}
      \hfill
      \caption{Streamlines of the average velocity field and contours of time averaged turbulent viscosity. }
\label{fig6_1}
\end{figure}

\begin{figure}[b]
    \centering
    \includegraphics[width=0.5\textwidth]{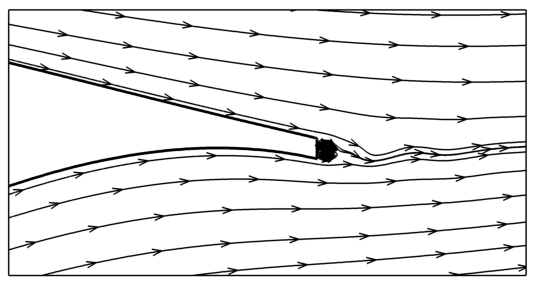}
    \caption{Trailing edge vortical structures observed at $z/R=0.6$}
    \label{fig:trl-edge}
\end{figure}

To further understand the flow evolution around each airfoil cut along the WT blade, from the 24 airfoils simulated, we have chosen 12 of them to present the time averaged streamlines pattern as well as the turbulent viscosity field. This information is presented in \Cref{fig6_1}.
The respective locations of the twelve chosen airfoils are identified by their distance from the root, represented by $z/R= 0.1, 0.15, 0.2, 0.25, 0.3, 0.35, 0.4, 0.5, 0.6, 0.7, 0.8, 0.9$. As we move from the root to the tip of the blade, we observe a distinctive pattern. Nearby the root at $z/R=0.1$, where the AoA is particularly large, there exists an early separation of the boundary layer, generating a large vortical structure that turns clockwise. The interaction of this primary vortical structure with the shear flow generated at the airfoil trailing edge gives birth to a second smaller vortical structure which turns counterclockwise and appears downstream of the first one. The two vortical structures are observed until $z/R=0.35$ and they become smaller as we move towards the blade tip. In fact this effect is to be expected since the AoA decreases at a higher blade radius and as a result the boundary layer separates further downstream. At values of $z/R=0.5$ and higher, the boundary layer is mostly attached along the entire chord, minor separations are observed at the airfoil trailing edge. \Cref{fig:trl-edge} presents the time averaged streamlines at the airfoil trailing edge for $z/R=0.6$. 

If we analyze now the information presented in \Cref{fig6_1,fig:trl-edge} and linking it with that of \Cref{fig5_1}, it can be concluded that large vortical structures are generated when the boundary layer has an early separation and has associated large peak to peak amplitudes of the dynamic lift and drag forces. A delay of the separation of the boundary layer generates smaller vortical structures with smaller peak to peak lift and drag amplitudes. As soon as the boundary layer is attached along the entire chord, and due to the particular trailing edge shape of the studied airfoils, just very small vortical structures are observed in this location. The associated lift and drag dynamic forces present very small peak-to-peak amplitudes.

\begin{figure}[t]
    \centering
    \begin{subfigure}[b]{0.48\textwidth}
    \centering
    \includegraphics[width=\textwidth]{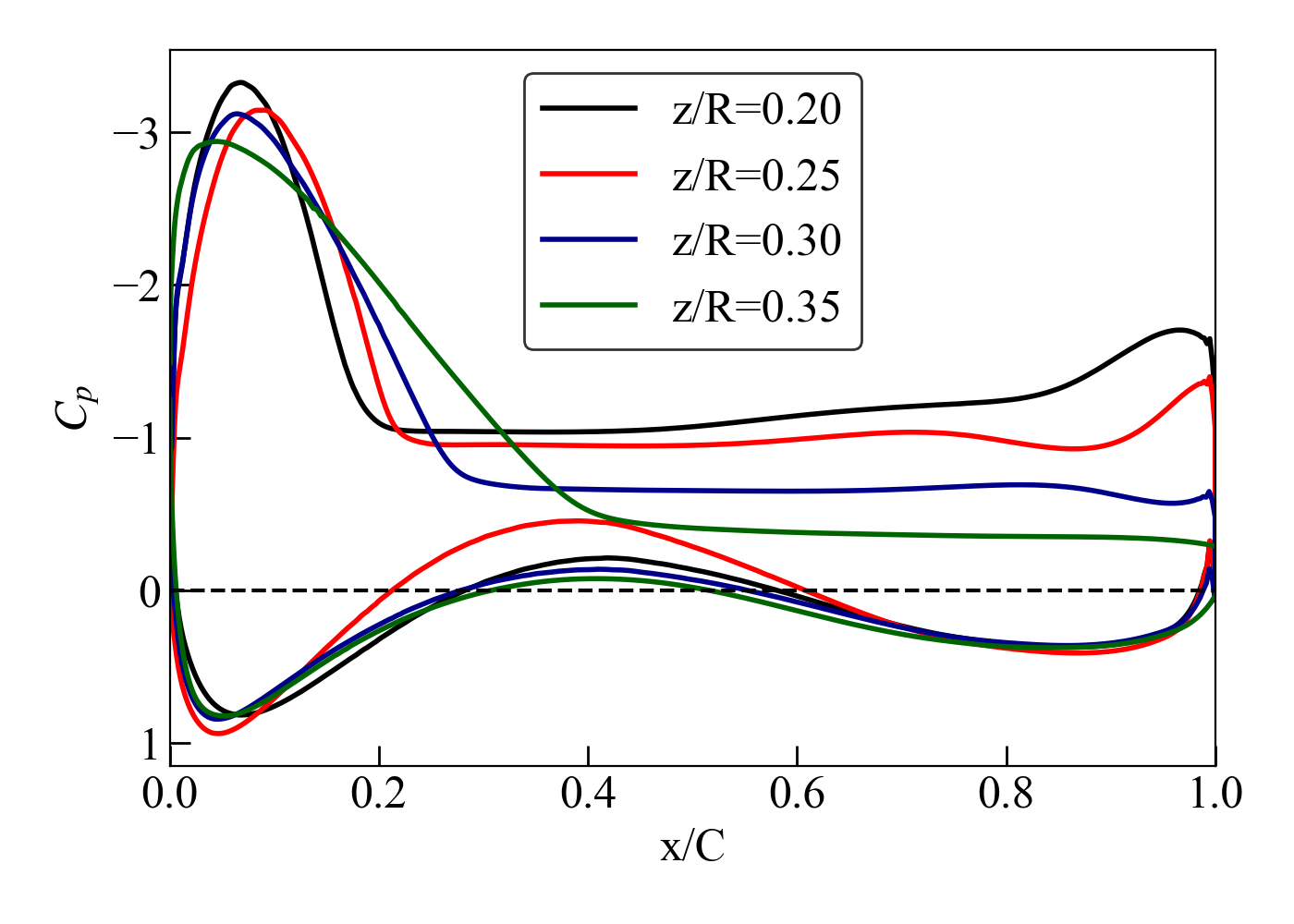}
    \caption{ }
    \end{subfigure}
    \begin{subfigure}[b]{0.49\textwidth}
    \centering
    \includegraphics[width=\textwidth]{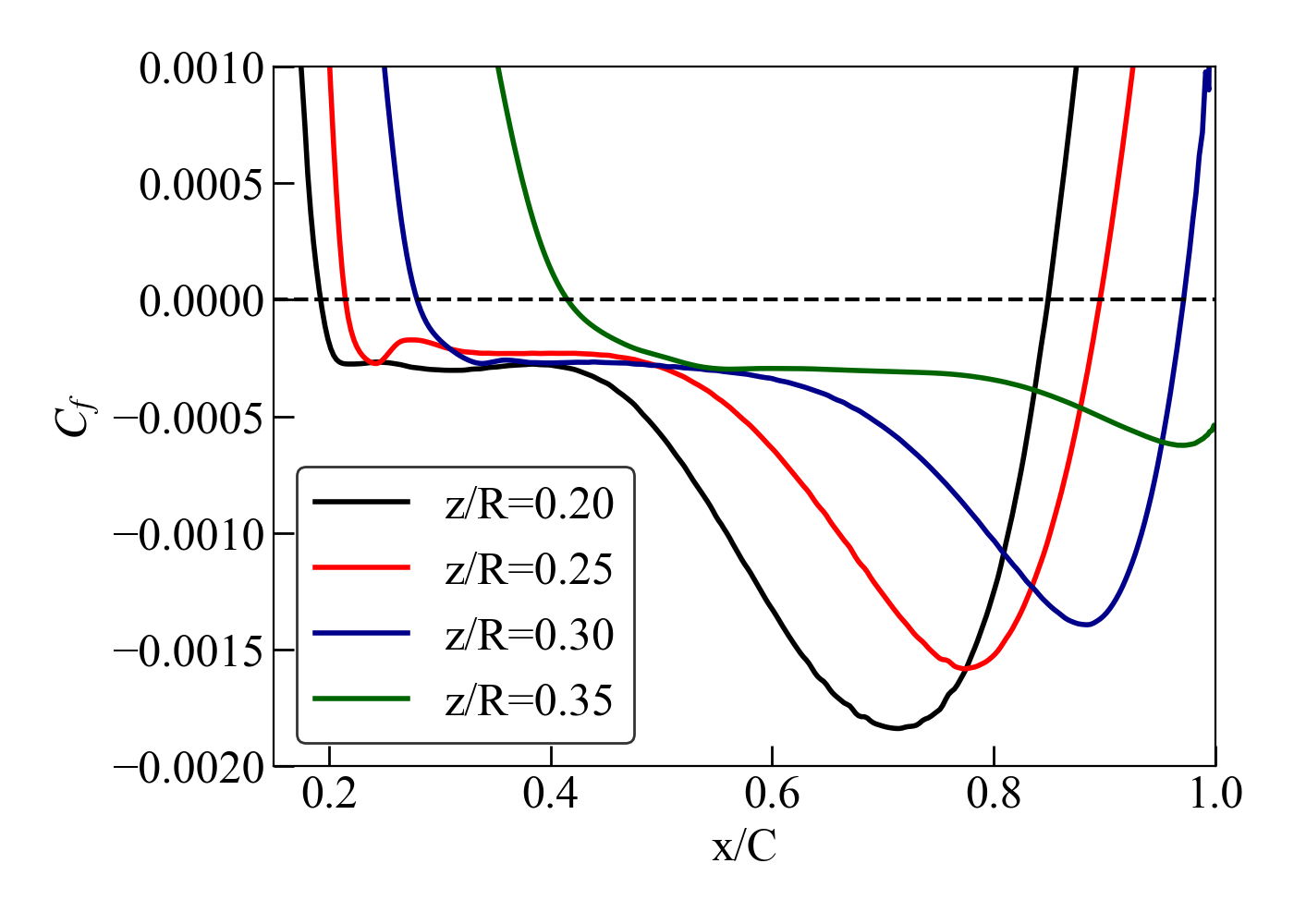}
    \caption{ }
    \end{subfigure}
    \caption{Distribution of pressure (a) and skin friction coefficient on the pressure surface (b) for baseline cases at four different cuts of blade.}
    \label{fig8}
\end{figure}

In order to gain inside of the exact pressure and shear stress distribution around each airfoil, as well as to obtain the exact location where the boundary layer separates, the pressure coefficient and skin friction coefficient distributions were obtained. \Cref{fig8} highlights these coefficients for the following four chosen airfoils, $z/R=0.2, 0.25, 0.3, 0.35$. 
Focusing on the skin friction coefficient, it is worth noting that at the boundary layer separation point, the shear stresses become zero and their values become negative further downstream. Notice as well that as the airfoil is located at a higher radius, the boundary layer separates further downstream, their respective values for $z/R=0.20, 0.25, 0.30, 0.35$, are $x_s/C=0.22, 0.24, 0.27$ and $0.41$, see \Cref{tab_all_sections} in the appendix.  
When examining the pressure coefficient on the upper surface of each airfoil, we noticed a correlation with the size of the separated boundary layer. 
As the boundary layer thickness decreases (the boundary layer separates further downstream), the maximum negative value of the pressure coefficient also reduces.
In fact, the pressure coefficient remains quite stable along the chord length after the point where the boundary layer separates, tending towards the atmospheric pressure as the separation point moves downstream. A delay of the boundary layer separation has been associated with a smaller zone of negative skin friction values, indicating a more favorable aerodynamic condition. In summary, the trends observed in \Cref{fig8} align very well with what is observed in \Cref{fig6_1} and demonstrate a direct relationship between the location of the separated boundary layer, the pressure coefficient on the airfoil upper surface and the skin friction coefficient. Delaying the separation of the boundary layer leads to an increase of the pressure below the airfoil and reduces the negative skin friction values, enhancing the aerodynamic performance.

\begin{figure}[t]
    \centering
    \includegraphics[width=0.9\textwidth]{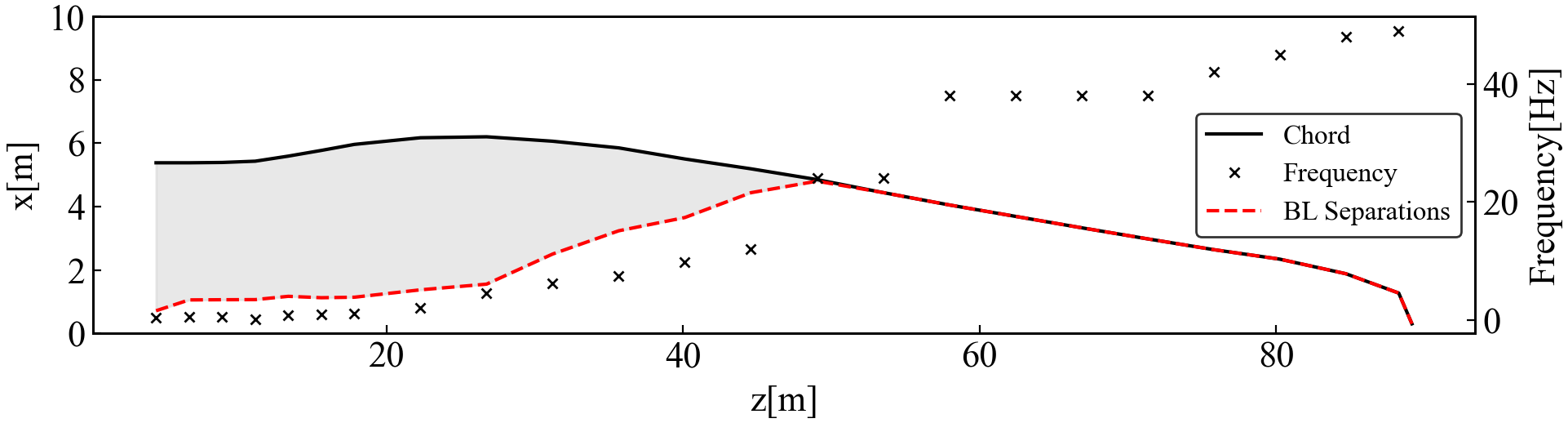}
    \caption{Chord length, boundary layer separation point, and its associated vortex shedding frequency, for each of the 24 airfoils evaluated in this study.}
    \label{fig:sep}
\end{figure}

\Cref{fig:sep} introduces the boundary layer evolution and its associated frequency along the entire wind turbine blade. The solid black line characterizes the shape of the blade along its length (from root to tip), which it is called the chord. The 'x' marks indicate the vortex shedding main frequency associated with each airfoil. The red dashed line shows the boundary layer separation point along the chord and for each airfoil as we move from the root to the tip of the blade.
As previously observed in \Cref{fig6_1}, the boundary layer separation begins at around 13\% of the chord length for $z/R=0.05$. As we move towards the tip, the separation point gets closer to the trailing edge of the corresponding airfoil. For instance, when z/R = 0.5, the separation occurs at around 92\% of the chord length. 
Early boundary layer separation locations have associated large vortical structures and small vortex shedding frequencies.
Near the root ($z/R = 0.05$), the frequency is low, around 0.23 Hertz. However, as we move towards the blade tip and the separation point gets closer to the trailing edge, the frequency keeps increasing. For blade locations higher than $z/R=0.5$, the boundary layer remains attached to each airfoil, and there's a sudden jump in the vortex shedding frequency. This is due to the drastic reduction of the dimension of the vortical structures generated at the airfoil trailing edge, see \Cref{fig:trl-edge}. In the upper part of the blade where the boundary layer is attached, the increase of frequency as we move towards the blade tip is linked to the increase of the relative velocity associated to each airfoil. Higher relative velocities involve higher flow kinetic energies, smaller vortical structures at the trailing edge and higher frequencies.

\section{AFC Parameter Optimization}

\subsection{Approaches to Optimization}
Based on the author's previous research \cite{tousi2021active}, we decided to use a 
multi-objective optimization to obtain the five AFC parameters that maximize lift and efficiency on each airfoil. 

The multi-objective approach chosen will allow the identification of the entire Pareto front in one go and, consequently, determine the best solutions considering the complete set of objective functions. A multi-objective optimization approach shall be defined in its minimization form as: 

\begin{equation}
     \left\{\begin{array}{lll}
        min & (f_1(\overrightarrow{x}),f_2(\overrightarrow{x})...f_k(\overrightarrow{x}) ), & K=1, 2,...,K\\
            & subject \:  to: & \\
        g_l &  (\overrightarrow{x})\geq 0, & l=1, 2, ..., L\\
        h_m &  (\overrightarrow{x})=0, & m=1, 2, ..., M\\
            & Where: &\\
        \overrightarrow{x}   &  \in X, & X \: feasible \: set \: of \: decision \: variables.\\
    \end{array}\right.
\end{equation}

To make things clearer, let's define some terms:
\begin{itemize}
    \item Objective Functions (FOs): These are the measures of how good our choices are(fitness functions). They tell us how well our design variables $f_1(\overrightarrow{x}),f_2(\overrightarrow{x})...f_k(\overrightarrow{x})$ are working toward our goals. Being the K the number of objective functions.
\end{itemize}
\begin{itemize}
    \item Design Variables (DVs) refers to the array of decision variables, which are the independent variables within the optimization problem, denoted as $\overrightarrow{x}\in X$.
\end{itemize}
\begin{itemize}
    \item Constraints (Ctrs): includes a collection of conditions that must be met for the Decision Variables (DVs) to be deemed feasible. These constraints consist of both inequality expressions ($g_l$) and equality expressions ($h_m$), where L represents the number of inequality constraints and M represents the number of equality constraints.
\end{itemize}

In our study, we are using something called a Genetic Algorithm (GA). Genetic Algorithms were first suggested by Holland \cite{holland1992adaptation} and later improved by Koza and Goldberg \cite{koza1994genetic}. They work by mimicking how nature evolves, selecting the best solutions, and combining them to create even better ones. The reason behind this decision is because a GA is good at finding the best solutions when we have multiple goals, which we call the Pareto front. It is a robust methodology capable of handling configurations that cannot be computed due to mesh generation or any process related to the numerical evaluation of each case.
Our Genetic Algorithm, outlined in Algorithm 1, is coupled with a Python script that performs the steps required to change the shape of the design, the boundary conditions, and the mesh, as a function of the design variables considered. And executes the CFD simulations. 

\begin{algorithm}
    \SetKwFunction{EvaluateObjectives}{EvaluateObjectives}
    \SetKwFunction{PerformCFDCalculations}{PerformCFDCalculations}
    \SetKwFunction{InitializePopulation}{InitializePopulation}
    \SetKwFunction{ApplySelection}{ApplySelection}
    \SetKwFunction{ApplyCrossover}{ApplyCrossover}
    \SetKwFunction{ApplyMutation}{ApplyMutation}
    \SetKwFunction{AdjustBoundaryConditions}{AdjustBoundaryConditions}
    \SetKwFunction{UpdateGeometry}{UpdateGeometry}
    \SetKwFunction{GenerateMesh}{GenerateMesh}
    \SetKwFunction{PostProcessing}{PostProcessing}
    
    \InitializePopulation{}\;
    \While{Not meeting termination criteria}{
        \EvaluateObjectives{}\;
            \:\:\:\AdjustBoundaryConditions{}\;
            \:\:\:\UpdateGeometry{}\;
            \:\:\:\GenerateMesh{}\;
            \:\:\:\PerformCFDCalculations{}\;
            \:\:\:\PostProcessing{}\;
        \ApplySelection{}\;
        \ApplyCrossover{}\;
        \ApplyMutation{}\;
    }
    \caption{Genetic Algorithm and Python script}\label{alg:modified_algorithm}
\end{algorithm}

To begin with, the random generation of a population takes place in the \texttt{InitializePopulation()} function. A population size of 20 individuals is chosen, striking a balance between method performance and computational cost. Subsequently, the \texttt{EvaluateObjectives()} function processes and computes the population by determining the objective functions for each individual. An in-house Python script facilitates the preparation of each individual by linking the mesh generator (GMSH), the CFD package (OpenFoam), and the GA. This script automates the translation of formats across different packages, modifies the mesh, and applies boundary conditions. After computing the population, optimization operators—\texttt{ApplySelection()}, \texttt{ApplyCrossover()}, and \texttt{ApplyMutation()}—come into play. In the present study's implementation, the \texttt{ApplySelection()} operator adopts a $\mu + \lambda$ strategy with a Crowded-Comparison Operator\cite{deb2001constrained}. The \texttt{ApplyCrossover()} operator utilizes a Simulated Binary Crossover (SBX)\cite{deb1995simulated}, and the \texttt{ApplyMutation()} operator employs a Polynomial Mutation\cite{deb2015multi}. Importantly, the optimizer keeps track of individuals that survive to the next generation without altering their design variables, avoiding the re-computation of their objective functions. In resource-intensive and time-consuming applications, understanding this behavior is crucial to prevent redundant CFD analyses. The parameters of the GA used in this study are detailed in \Cref{table04}.

\begin{table}[t]
\centering
\caption{Genetic Algorithm Parameters}
\begin{tabular}{lr} 
\hline
Parameter & Value \\
\hline
Population Size & 20 \\
Crossover Probability & 0.9  \\
Mutation Probability & 0.1  \\
\hline
\end{tabular}
\label{table04}
\end{table}

\subsection{Defining Objective Functions and Design Variables}

In this study, we seek to enhance the lift coefficient ($C_l$) and the aerodynamic efficiency ($\eta=C_l/C_d$) of several airfoils cut along the WT blade. The Genetic Algorithm (GA) employed approaches to the optimization task as a minimization problem, aiming to minimize the objective functions. Considering this approach, we defined two objective functions, $f_1$ and $f_2$. These functions are minimized when both the lift coefficient and efficiency reach their maximum values.

\begin{equation}
    f_1=  -C_L
\end{equation}

\begin{equation}
    f_2=  -\eta
\end{equation}

\begin{figure}[t]
     \centering
     \begin{subfigure}[b]{0.45\textwidth}
         \centering
         \includegraphics[width=\textwidth]{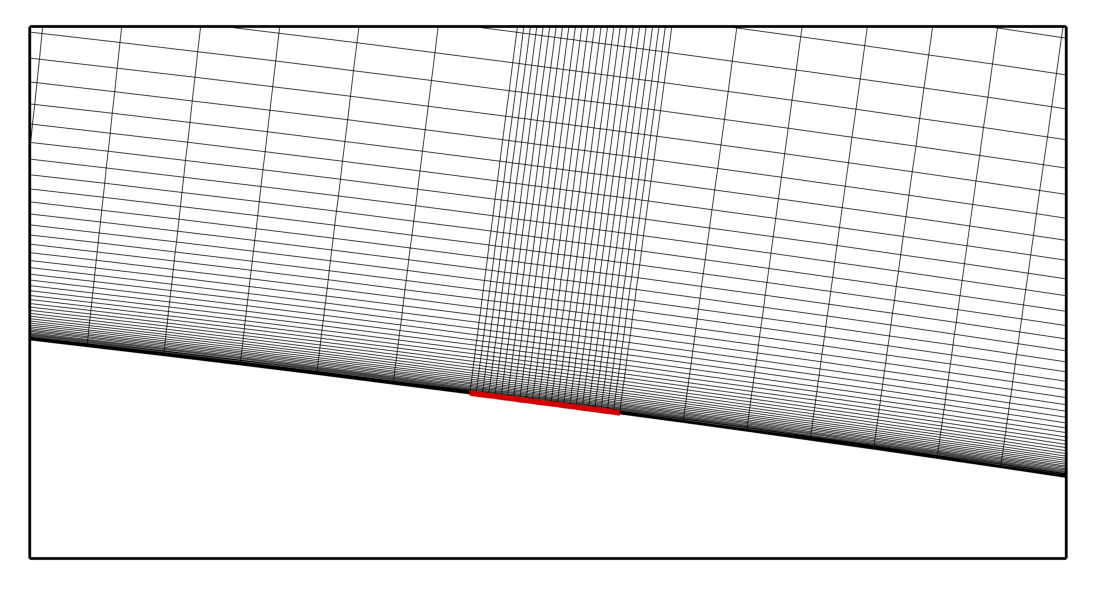}
     \end{subfigure}
    \begin{subfigure}[b]{0.45\textwidth}
         \centering
         \includegraphics[width=\textwidth]{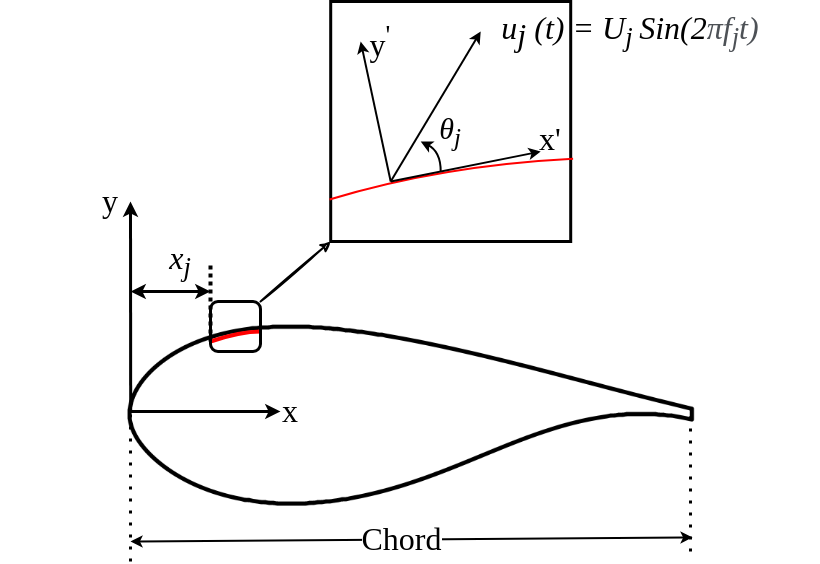}
     \end{subfigure}
     \hfill
     \caption{a) Mesh view near the jet automatically generated for $z/R=0.35$(Red line shows AFC jet edge). b)General view of the jet location and its inclination angle.}
\label{figure-jet}
\end{figure}

The optimization process was applied to three blade sections at $z/R=0.25$, $0.30$, and $0.35$, involving three independent optimization analyses. Design variables were defined as five parameters related to the synthetic jet, namely non-dimensional frequency ($F^+$), momentum coefficient ($C_\mu$), jet angle ($\theta$), jet position ($x/C$), and jet width ($h/C$). \Cref{figure-jet} illustrates a schematic representation of the geometric design parameters, zooming around the jet position for one of the automatically generated meshes.

In Algorithm~\ref{alg:modified_algorithm} \texttt{AdjustBoundaryConditions()} function, the first three design variables ($F^+$, $C_\mu$, and $\theta$) were chosen. Following this, the \texttt{UpdateGeometry()} function, based on the last two design variables ($x/C$ and $h/C$), generated a new geometry using the GMSH program. After applying these geometrical changes, the \texttt{GenerateMesh()} function created a new mesh, determining the required number of cells along the groove width to ensure a high mesh quality.
With the completion of these steps, the CFD simulations were ready to commence, facilitated by the\texttt{PerformCFDCalculations()} function. Finally, the two objective functions underwent post-processing in the \texttt{PostProcessing()} function, and the results were then transferred to the three Genetic Algorithm (GA) operators.

\begin{table}[t]
\centering
\caption{Active flow control design variable and their evaluation for the different blade's section}
\begin{tabular}{lllllll} 
\hline
Z/R & AoA & $F^+$ & $C_\mu$ & $\theta^\circ$ & $x_j/C$ & $h/C$\\
\hline
0.25  & $23.8^\circ$ & 0.1-10 & 0.0001-0.01 & 5-45 & $x_s/C \pm 0.15C$ & 0.005-0.015\\
0.30 & $20.
5^\circ$ & 0.1-10  & 0.0001-0.01 & 5-45 & $x_s/C \pm 0.15C$ & 0.005-0.015\\
0.35 & $17.49^\circ$ & 0.1-10  & 0.0001-0.01 & 5-45 & $x_s/C \pm 0.20C$ & 0.005-0.015\\
\hline
\end{tabular}
\label{table05}
\end{table}

\Cref{table05} presents the ranges considered for the five AFC design variables and for each of the three different sections analyzed. For all cases, the variables were evaluated within specific intervals: $F^+\in[0.1:10]$, $C_\mu \in[0.00001:0.01]$, $\theta \in [5^\circ:45^\circ]$, $x/C \in [x_s /C \pm 0.15C]$, and $h/C \in [0.005: 0.015]$. These ranges were chosen based on previous studies. For instance, pulsating flows commonly use frequencies around $F^+ = 1$. Since pulsating frequency significantly influences the boundary layer performance, we extended the limits to be ten times smaller and larger than this baseline value. The upper limit for momentum coefficients was determined by the maximum values used in experimental works by Gilarranz et al. \cite{gilarranz2005new} and Goodfellow et al. \cite{goodfellow2013momentum}. Regarding the jet width (h/C), a wide range was explored by the optimizer and our analysis suggested that the optimal solution often involved employing rather small widths. For the jet position (x/C), we ensured that the separation point at each section remained in the middle of the chosen margins to minimize computational costs.

\section{AFC results}

\begin{figure}[t!]
     \centering
     \begin{subfigure}[b]{0.49\textwidth}
         \centering
         \includegraphics[width=\textwidth]{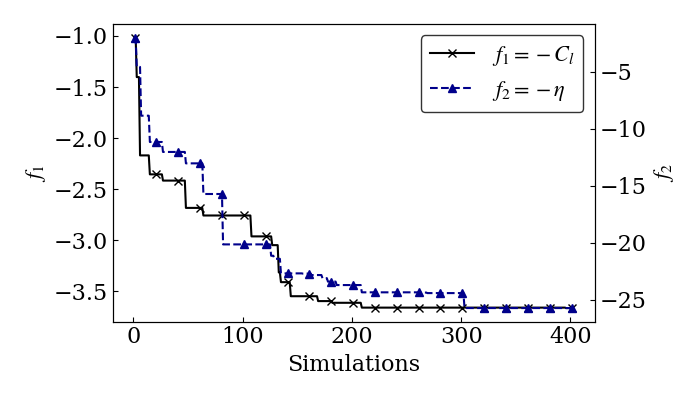}
         \caption{z/R=0.25}
         \label{11a}
     \end{subfigure}
     \begin{subfigure}[b]{0.49\textwidth}
         \centering
         \includegraphics[width=\textwidth]{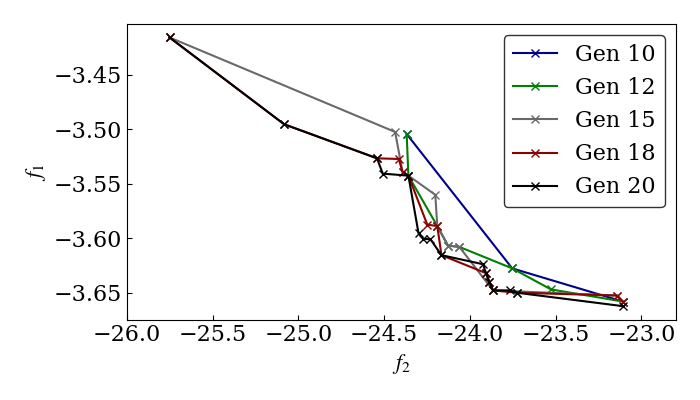}
         \caption{z/R=0.25}
         \label{11b}
     \end{subfigure}
     \begin{subfigure}[b]{0.49\textwidth}
         \centering
         \includegraphics[width=\textwidth]{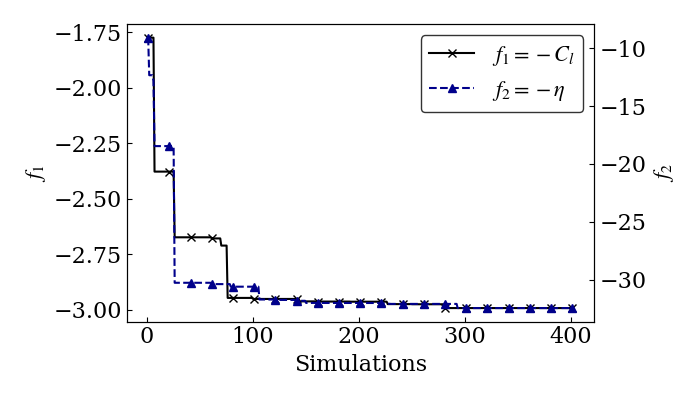}
         \caption{z/R=0.30}
         \label{11c}
     \end{subfigure}
    \begin{subfigure}[b]{0.49\textwidth}
         \centering
         \includegraphics[width=\textwidth]{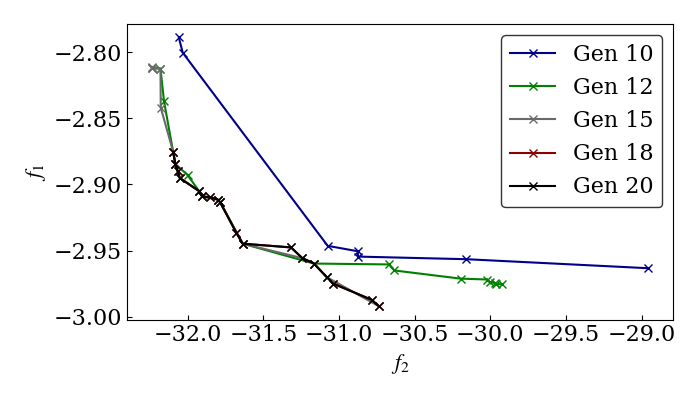}
         \caption{z/R=0.30}
         \label{11d}
     \end{subfigure}
    \begin{subfigure}[b]{0.49\textwidth}
         \centering
         \includegraphics[width=\textwidth]{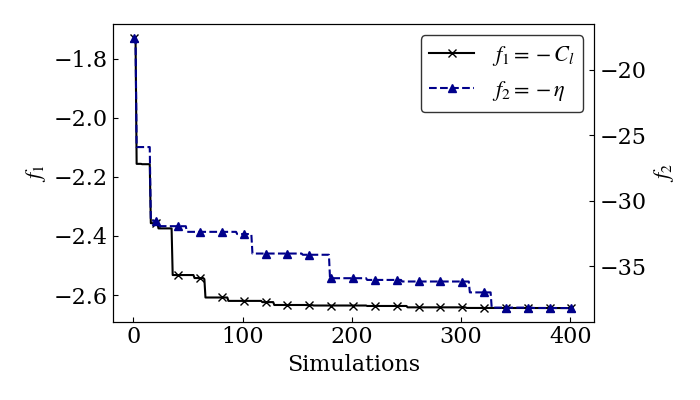}
         \caption{z/R=0.35}
         \label{11e}
     \end{subfigure}
    \begin{subfigure}[b]{0.49\textwidth}
         \centering
         \includegraphics[width=\textwidth]{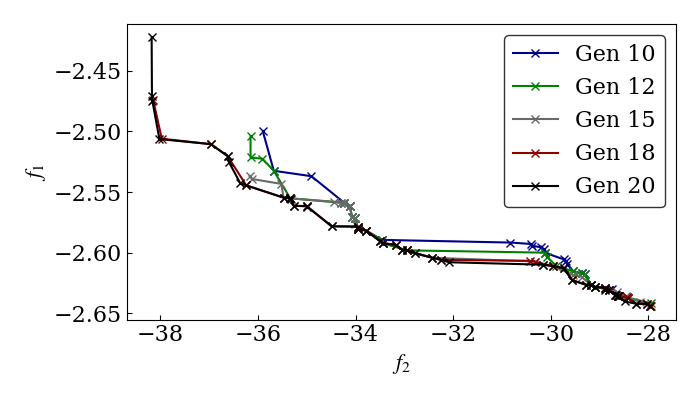}
         \caption{z/R=0.35}
         \label{11f}
     \end{subfigure}
     \caption{Evaluation of objective functions over number of simulations (a, c, and e) and Pareto front progression across generations (b, d, an f) .}
\label{convergency}
\end{figure}

The optimization process described herein was repeated until the specified termination conditions were satisfied. Initially, the termination criterion involved the computation of a minimum of 200 individuals, subsequently revised to 400 individuals. This adjustment was motivated by the findings elucidated in \Cref{convergency}, which provides insights into the convergence evaluation and Pareto front history across three distinct airfoils characterized by z/R ratios of 0.25, 0.30, and 0.35. 

In all situations, about 400 CFD simulations were needed to make the two objective functions ($f_1$) and ($f_2$) as small as possible. This means making the lift coefficient ($C_l$) and aerodynamic efficiency ($\eta$) as high as possible. In all cases, there was a minor difference in the optimized variables in the last 5 generations. Which means that after 300 simulations the five optimum AFC parameters remained almost constant and therefore both objective functions did not really show any improvement. After 400 simulations, there were only negligible variations in the objective functions. The same information is obtained when observing the Pareto fronts in \Cref{11b,11d,11f}. After 15 generations (groups of 20 simulations), the Pareto front curves presented minor variations.

\begin{figure}[]
     \centering
     \begin{subfigure}[b]{0.3\textwidth}
         \centering
         \includegraphics[width=\textwidth]{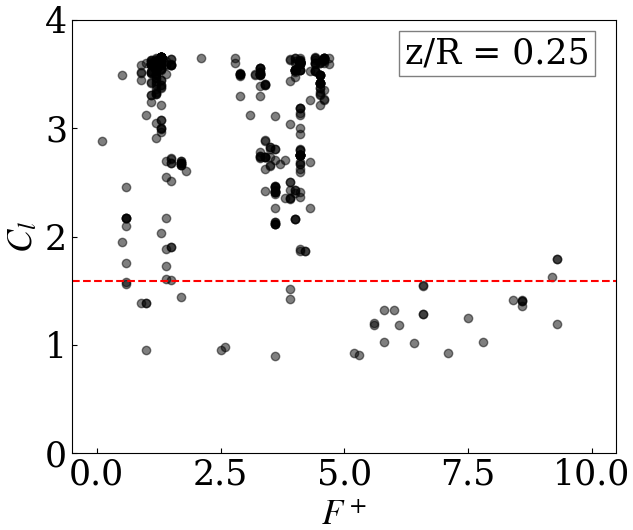}
     \end{subfigure}
     \begin{subfigure}[b]{0.3\textwidth}
         \centering
         \includegraphics[width=\textwidth]{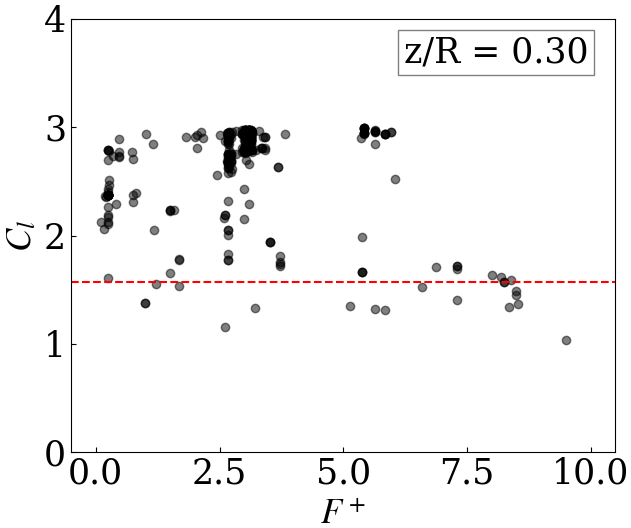}
     \end{subfigure}
     \begin{subfigure}[b]{0.3\textwidth}
         \centering
         \includegraphics[width=\textwidth]{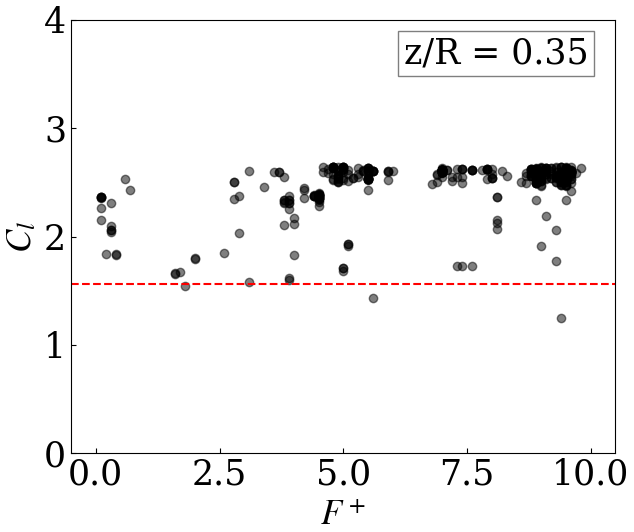}
     \end{subfigure}
     \begin{subfigure}[b]{0.3\textwidth}
         \centering
         \includegraphics[width=\textwidth]{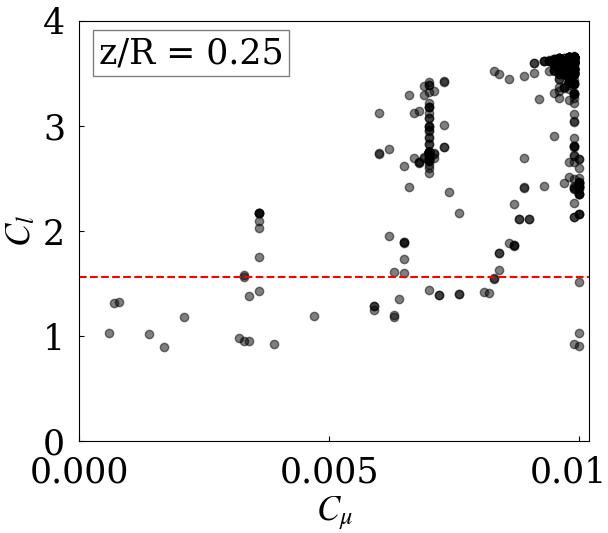}
     \end{subfigure}
     \begin{subfigure}[b]{0.3\textwidth}
         \centering
         \includegraphics[width=\textwidth]{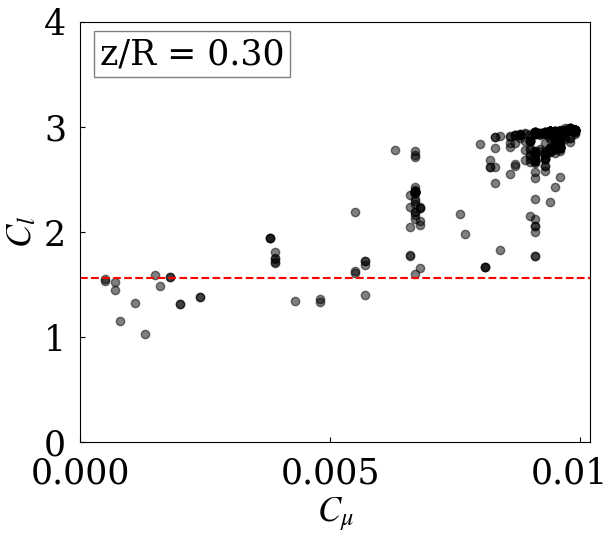}
     \end{subfigure}
     \begin{subfigure}[b]{0.3\textwidth}
         \centering
         \includegraphics[width=\textwidth]{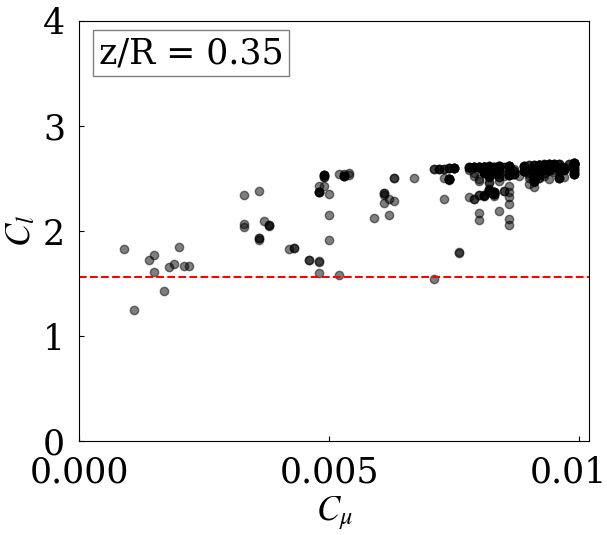}
     \end{subfigure}
     \begin{subfigure}[b]{0.3\textwidth}
         \centering
         \includegraphics[width=\textwidth]{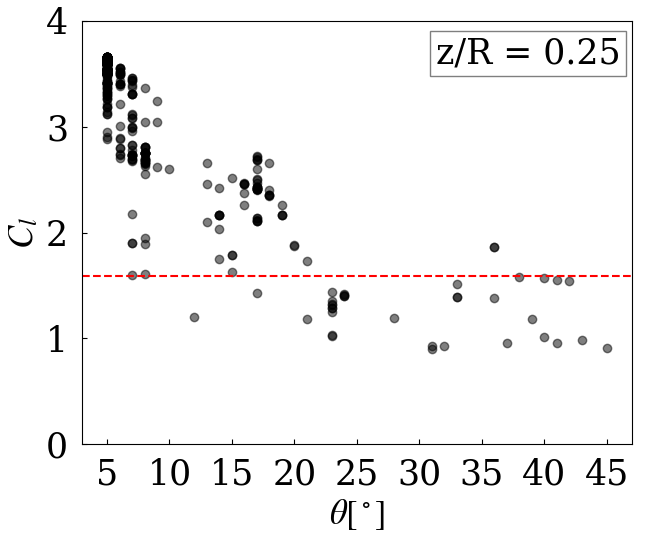}
     \end{subfigure}
     \begin{subfigure}[b]{0.3\textwidth}
         \centering
         \includegraphics[width=\textwidth]{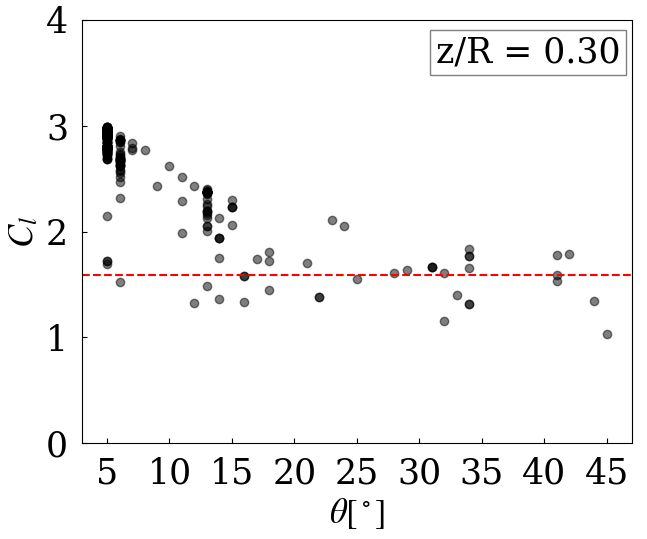}
     \end{subfigure}
     \begin{subfigure}[b]{0.3\textwidth}
         \centering
         \includegraphics[width=\textwidth]{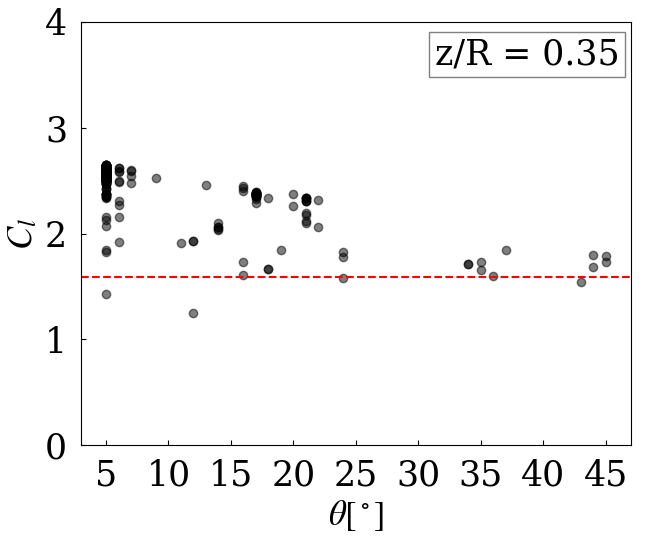}
     \end{subfigure}
     \begin{subfigure}[b]{0.3\textwidth}
         \centering
         \includegraphics[width=\textwidth]{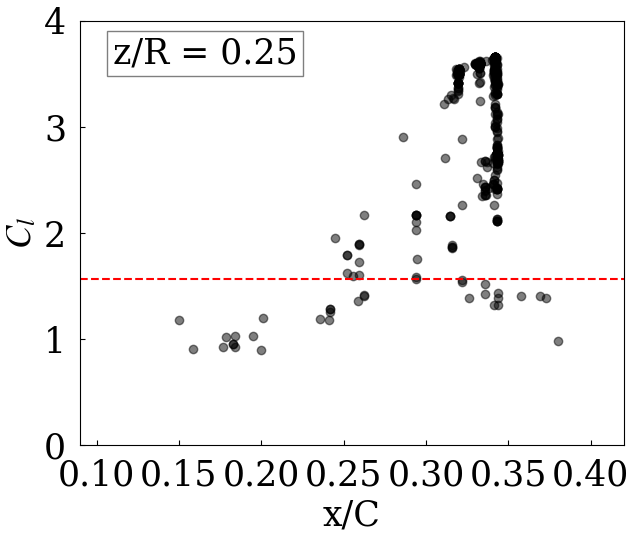}
     \end{subfigure}
     \begin{subfigure}[b]{0.3\textwidth}
         \centering
         \includegraphics[width=\textwidth]{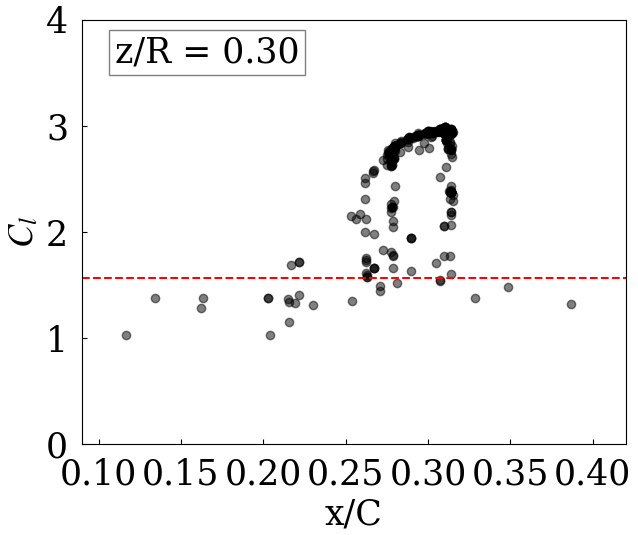}
     \end{subfigure}
     \begin{subfigure}[b]{0.3\textwidth}
         \centering
         \includegraphics[width=\textwidth]{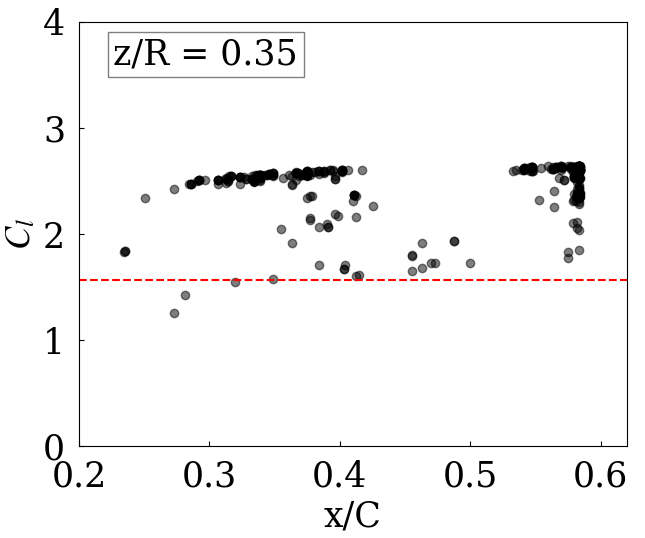}
     \end{subfigure}
     \begin{subfigure}[b]{0.3\textwidth}
         \centering
         \includegraphics[width=\textwidth]{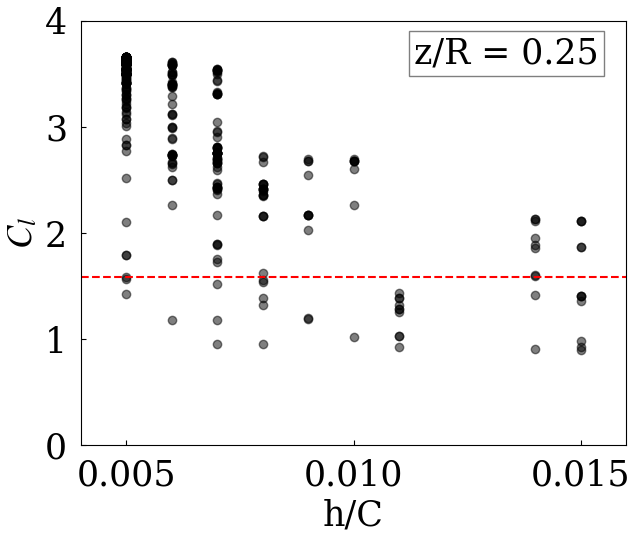}
     \end{subfigure}
     \begin{subfigure}[b]{0.3\textwidth}
         \centering
         \includegraphics[width=\textwidth]{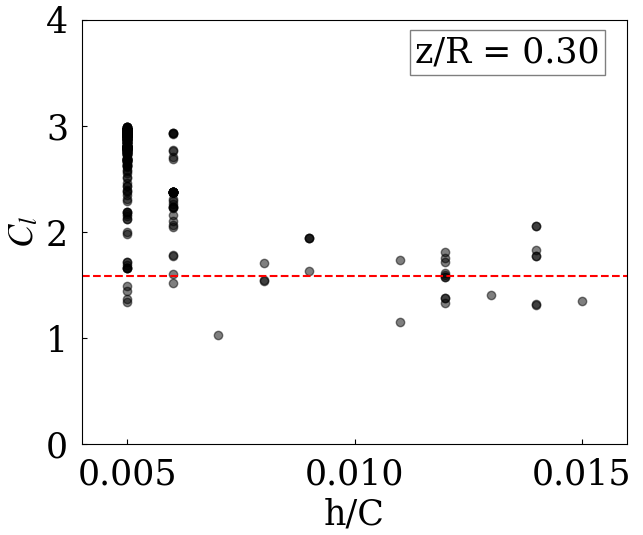}
     \end{subfigure}
     \begin{subfigure}[b]{0.3\textwidth}
         \centering
         \includegraphics[width=\textwidth]{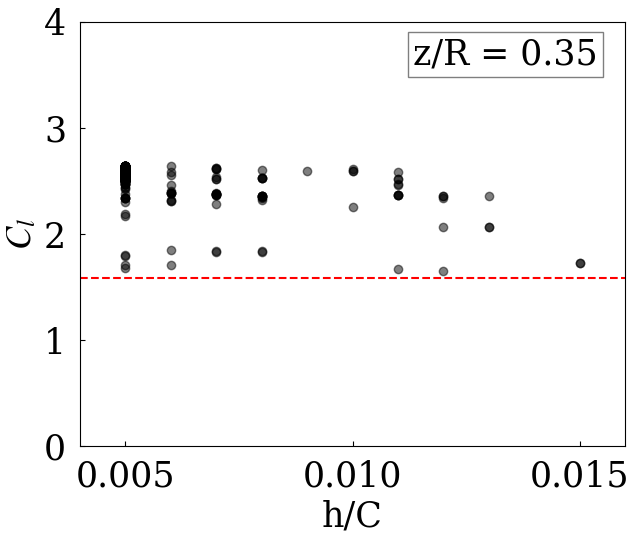}
     \end{subfigure}
     \caption{Lif coefficient $C_l$ versus non-dimensional AFC parameters and for all the CFD cases analyzed. Each column represents an airfoil.}
\label{all-opt}
\end{figure}

The primary objective of investigating the optimal Active Flow Control (AFC) parameters across various sections of the blade, 
is to assess the evolution of each AFC parameter along the blade. This evaluation aims to completely reattach the flow or delay the boundary layer separation along the entire blade.
\Cref{all-opt} illustrates the lift coefficient for each of the 400 Computational Fluid Dynamic (CFD) cases studied and as a function of each AFC parameter. Each column delineates, for a given airfoil, the lift coefficient in relation to the different AFC parameters. From left to right, the airfoils evaluated in each column are z/R= 0.25, 0.30, and 0.35, respectively.
The value of the lift coefficient, contingent upon a specific AFC parameter, across diverse sections of the blade, is depicted within each row of \Cref{all-opt}.  For reference, a horizontal red line across all graphs indicates the time averaged baseline lift coefficient for each blade section. 

From the first row, it is seen that the optimum non-dimensional jet pulsating frequency increases with the $z/R$ increase. The second row clarifies that, regardless of the airfoil considered, the optimum momentum coefficient falls nearby the maximum value pre-established in the simulations, suggesting that enlarging the upper limit of the pre-established $C_\mu$ should be considered in future optimizations. The optimum injection angle, which is introduced in the third row of the figure, appears to be for all three airfoils, the minimum one considered, therefore suggesting smaller jet inclination angles should be evaluated in future optimizations. A similar trend is observed when analyzing the optimum groove width $h/C$. For all cases, the minimum groove width evaluated appears to be the optimum one, therefore opening a door to studying even smaller groove widths in the future. Groove positions $x/C$ on the other hand, do suffer a considerable change as a function of the airfoil considered. The position moves further downstream as $z/R$ grows.

\begin{table}[t]
    \caption{Optimum AFC parameters obtained for the maximum lift and maximum efficiency cases and for each of the three airfoils considered z/R=0.25, 0.30, and 0.35.}
    \centering
    \begin{tabular}{cllllllllllll}
    \hline
     $z/R$ & Cases & $F^+$ & $C_\mu$ & $\theta^\circ$ & $x_j/C$ & $h/C$ & $C_L$ & $C_L\%$ & $\eta$ & $\eta\%$ \\
    \hline
    0.25  & Max lift & 4.4 & 0.0099 & 5 & 0.34 & 0.005 & 3.66 & 130.6 & 23.71 & 540 \\
        & Max efficiency & 4.5 & 0.0098 & 5 & 0.31 & 0.005 & 3.41 & 115.09 & 25.75 & 594.88 \\
        & Baseline &-&-&-&-&-& 1.58 &-& 3.70 &-\\
            \cline{2-11}
    0.30 & Max lift & 5.41  &  0.0098  &  5  &  0.32  &  0.005  &  2.99  &  90.44  &  30.73  &  428.00 \\
        & Max efficiency & 5.64  &  0.0093  &  5  &  0.29  &  0.005  &  2.84  &  80.89  &  32.47  &  457.90 \\
        & Baseline &-&-&-&-&-&1.57&-&5.82&-\\
            \cline{2-11}
    0.35  & Max lift & 9.51 &  0.0093  &  5  &  0.58  &  0.005  &  2.63  &  68.94  &  28.43  &  100.52 \\
        & Max efficiency & 9.73  &  0.0074  &  5  &  0.33  &  0.005  &  2.49  &  60.23  &  35.90  &  153.18 \\
        & Baseline &-&-&-&-&-&1.56&-&14.18&-\\
    \hline
    \end{tabular}
    \label{table07}
\end{table}

As a summary of the results presented in \Cref{all-opt}, the optimized lift coefficients and efficiency values as well as the associated optimized AFC parameters, for the two optimized cases and for each airfoil evaluated are documented in \Cref{table07}. 
The first row introduces the definition of the blade section, the case considered, the five AFC parameters, the airfoil lift and efficiency, and their variation in percentage versus the baseline values.  The corresponding value of each parameter is given in the following rows and for each of the three airfoils evaluated. The first thing to notice is that, for a given airfoil, the optimum AFC parameters for the maximum lift and maximum efficiency are almost the same. According to the optimization findings, for section $z/R=0.25$ where the natural vortex shedding frequency is around $2.1 Hz$, the maximum lift is achieved with a non-dimensional jet pulsation frequency of $F^+ = 4.4$. For section $z/R=0.30$, where the natural vortex shedding frequency is 3.5 Hz, the maximum lift is attained when the non-dimensional pulsation frequency is of $F^+ = 5.41$. Lastly, at $z/R=0.35$, where the boundary layer separates further downstream compared to the previous sections, and the associated vortex shedding frequency is approximately of $7 Hz$, the maximum lift is reached when the jet non-dimensional pulsation frequency is of $F^+ = 9.5$. The tendency reads the higher the vortex shedding frequency the nearer to this value needs to be the jet pulsating frequency.
The optimum momentum coefficient is around $C_{\mu}=0.009$ for all cases studied. A constant value is also observed for the jet inclination angle $\theta$ and the groove width $h/C$, which for all the cases studied, remain constant and equal to the smallest values considered. This indicates that the jet injection/suction should be done almost tangentially to the surface, and small groove widths are the most appropriate, indicating that the SJA is more effective when concentrated in a particular small location. Regarding the jet position $x_j/C$, it is observed that under all cases studied, the optimum position is slightly downstream of the separation point $x_s/C$, see the values in the Appendix. Finally, it is observed that as $z/R$ increases, the lift coefficient and efficiency increase in percentage versus the baseline case decreases. In reality, this tendency is obvious, then as $z/R$ increases the boundary layer separates further downstream, leaving less room for improvement.   

\begin{figure}[b]
     \centering
     \begin{subfigure}[b]{0.32\textwidth}
         \centering      \includegraphics[width=\textwidth]{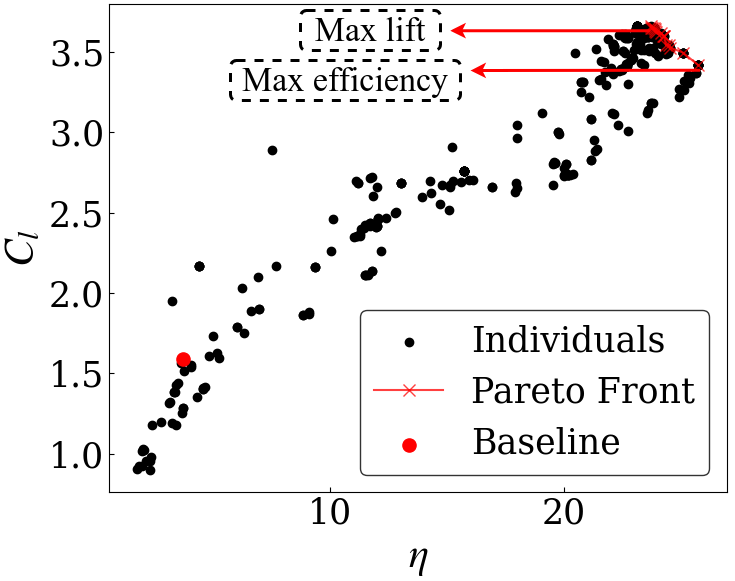}
     \caption{$z/R=0.25$}
     \end{subfigure}
     \begin{subfigure}[b]{0.32\textwidth}
    \centering
    \includegraphics[width=\textwidth]{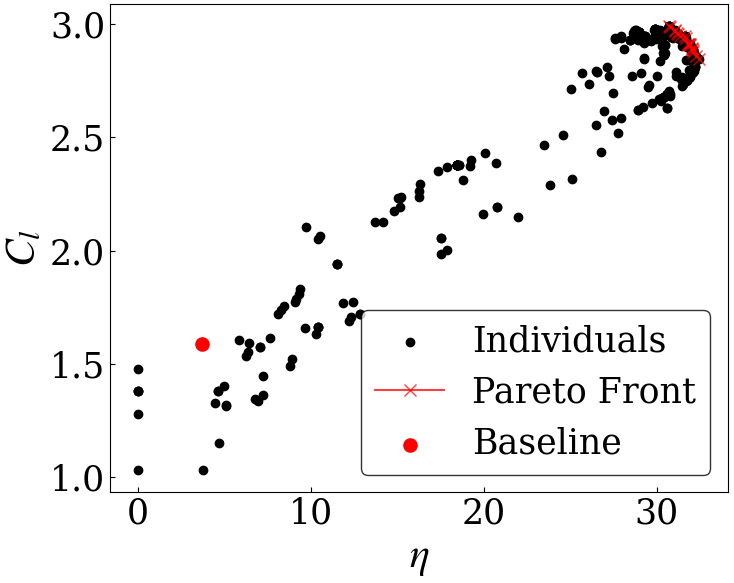}
     \caption{$z/R=0.30$}
     \end{subfigure}
     \begin{subfigure}[b]{0.32\textwidth}
    \centering
    \includegraphics[width=\textwidth]{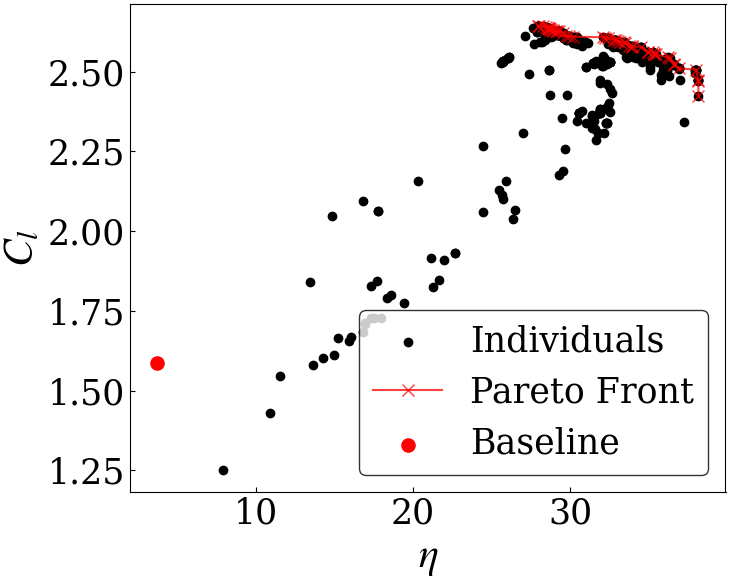}
     \caption{$z/R=0.35$}
     \end{subfigure}
     \caption{Lift coefficient versus airfoil efficiency for each individual, comparison with the baseline values.}
\label{fig-paret}
\end{figure}

In order to visualize the increase of lift coefficient and efficiency of each airfoil versus the respective baseline case values, \Cref{fig-paret} introduces the characterization of the two key objective functions explored in this study: aerodynamic efficiency and lift coefficient, across three distinct cases. Each figure presents over 400 data points derived from computational fluid dynamics (CFD) simulations. On the right-hand side of each graph, a red broken line denotes the Pareto front, indicating optimal values of airfoil parameters to achieve maximum efficiency, maximum lift, or an optimal compromise between both metrics. Additionally, a red dot within each figure represents the baseline characteristics of the airfoil for the respective case under examination.
\Cref{pareto-s42-all,pareto-s48-all,pareto-s54-all}, found in the Appendix, present a comprehensive summary of the five parameters corresponding to each point on the Pareto front for the three airfoils considered, as depicted in \Cref{fig-paret}. As a general trend, it is observed that as we move to a higher radius, higher $(z/R)$ values, the airfoils have associated smaller efficiencies increase versus the respective baseline case one. 
Progressing through the tables, we observe a decrease in the lift coefficient and an increase in aerodynamic efficiency, albeit without a glaring disparity between the two objectives. Notably, the top and bottom rows of each table denote the maximum lift and maximum efficiency cases, respectively. 
For $z/R=0.25$, the maximum lift has an aerodynamic efficiency associated of 23.71, as evidenced in \Cref{pareto-s42-all} of Appendix. Although not presented in the table, it is worth noting that the maximum lift case corresponds with the maximum torque generated by the airfoil. The optimal values for the five active flow control parameters contributing to the maximum aerodynamic lift coefficient are $F^+ = 4.4$, $C_\mu = 0.0099$, $\theta = 5^\circ$, $x_j/C = 0.34$, and  $h/C = 0.5\%$. 
Comparing the airfoil efficiency for the maximum lift case to the baseline case one, it reveals an impressive increase of almost $540\%$. Nevertheless, the maximum efficiency of the Pareto front is of $594.88\%$ although the corresponding torque is smaller than the one associated with the maximum lift case.
When examining the lift coefficients and airfoil efficiencies across all points on the Pareto front relative to the baseline case, improvements are evident in all optimal scenarios. 
Herein lies the significance of the Pareto front in this multi-objective optimization: it facilitates a nuanced discussion between the optimal outcomes for both objectives. 
Similar trends are observed in the other two sections under study, as detailed in \Cref{pareto-s48-all,pareto-s54-all}.

\begin{figure}[t!]
    \centering
    \begin{subfigure}[b]{0.4\textwidth}
    \includegraphics[width=\textwidth]{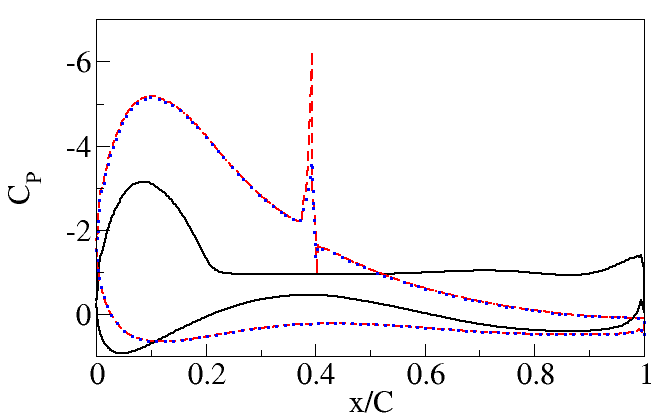}
    \caption{$z/R=0.25$}
    \end{subfigure}
    \begin{subfigure}[b]{0.4\textwidth}
    \includegraphics[width=\textwidth]{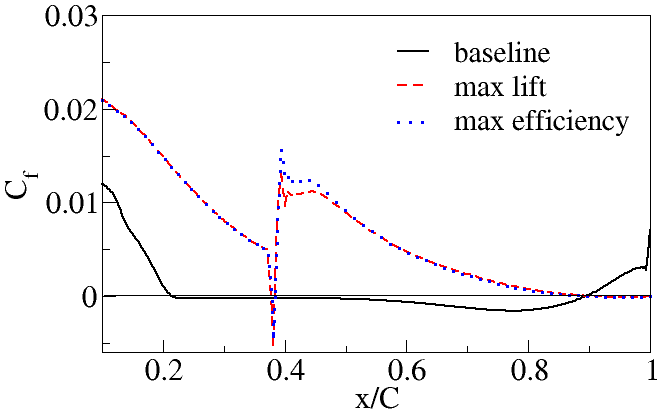}
    \caption{$z/R=0.25$}
    \end{subfigure}
    \begin{subfigure}[b]{0.4\textwidth}
    \includegraphics[width=\textwidth]{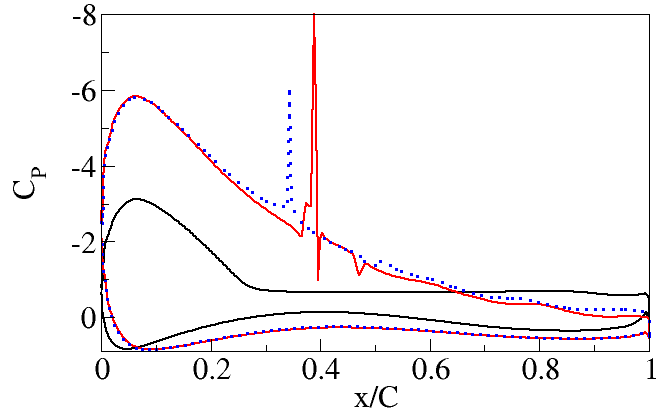}
    \caption{$z/R=0.30$}
    \end{subfigure}
    \begin{subfigure}[b]{0.4\textwidth}
    \includegraphics[width=\textwidth]{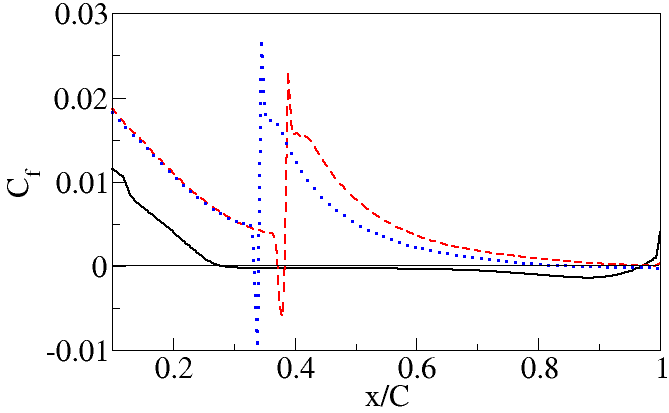}
    \caption{$z/R=0.30$ }
    \end{subfigure}
    \begin{subfigure}[b]{0.4\textwidth}
    \includegraphics[width=\textwidth]{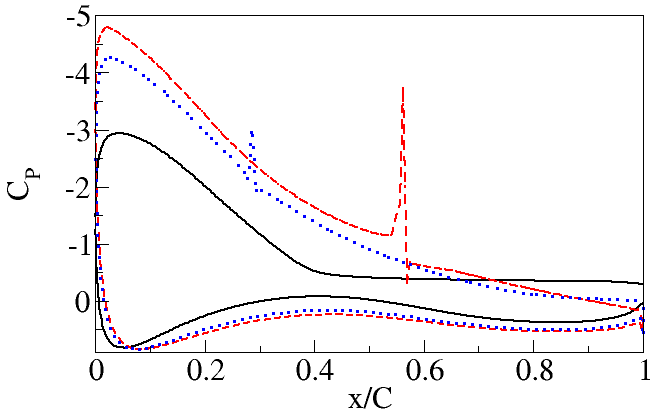}
    \caption{$z/R=0.35$ }
    \end{subfigure}
    \begin{subfigure}[b]{0.4\textwidth}
    \includegraphics[width=\textwidth]{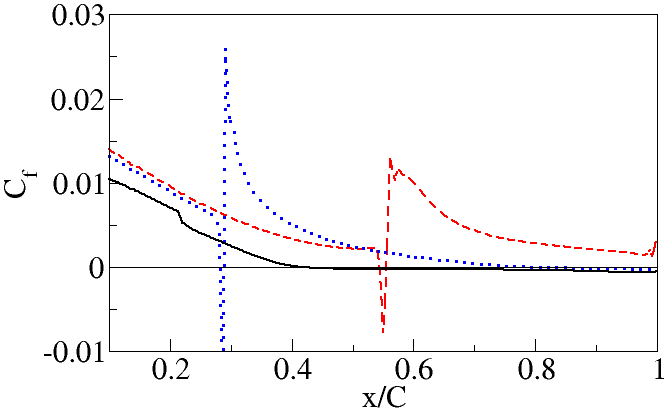}
    \caption{$z/R=0.35$ }
    \end{subfigure}
    \caption{Pressure $C_P$ and skin friction $C_f$ coefficients comparison for the baseline line, maximum lift, and maximum efficiency cases. (a), (b) $z/R=0.25$; (c), (d) $z/R=0.30$; (e), (f) $z/R=0.35$. }
    \label{fig14}
\end{figure}

Perhaps one of the best ways to visualize the effect of the AFC implementation on any airfoil is by comparing the pressure $C_P$ and friction $C_f$ coefficients obtained after the AFC implementation with the corresponding baseline case one. \Cref{fig14} illustrates the pressure and skin friction coefficients for each of the three optimizations studied, plotted as a function of chord length. It also includes a comparison between the baseline, maximum efficiency, and maximum lift cases. Looking at the curves for the pressure coefficient, irrespective of the case considered, it is evident that the AFC implementation increases/decreases the pressure below/above the airfoil. Pressure reduction is particularly notable between the leading edge and SJA locations.
The groove location for achieving maximum lift is always placed slightly downstream of the boundary layer separation point for the respective baseline case condition, see the friction coefficient curves. To obtain airfoil maximum efficiency and for the smallest $z/R$ ratio evaluated, the groove location is the same as for the maximum lift condition. But as the $z/R$ ratio increases, the optimum groove location to obtain maximum efficiency is displaced upstream versus the maximum lift one. In fact for the $z/R=0.35$ case, the groove location for maximum efficiency needs to be placed upstream of the baseline case boundary layer separation point. Detailed information on the groove locations and widths, as well as the rest of the AFC parameters, is to be found in \Cref{pareto-s42-all,pareto-s48-all,pareto-s54-all}.

\begin{figure}[t]
    \centering
    \begin{subfigure}[b]{0.7\textwidth}
    \includegraphics[width=\textwidth]{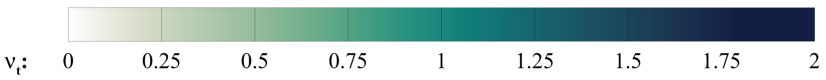}
    \end{subfigure}
    \begin{subfigure}[b]{0.3\textwidth}
    \includegraphics[width=\textwidth]{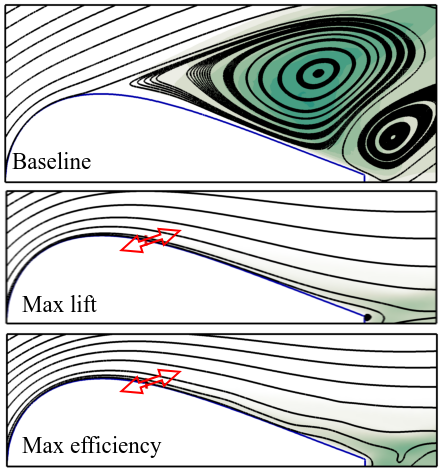}
    \caption{z/R=0.25}
    \end{subfigure}
    \begin{subfigure}[b]{0.303\textwidth}
    \includegraphics[width=\textwidth]{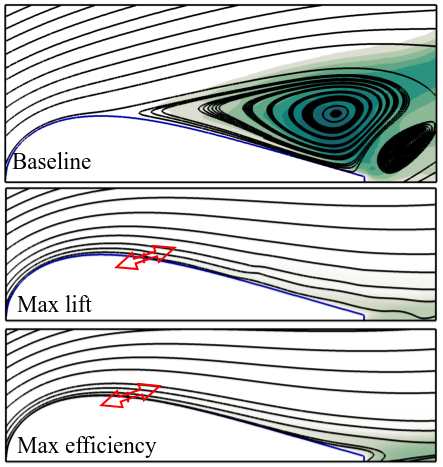}
    \caption{z/R=0.30}
    \end{subfigure}
    \begin{subfigure}[b]{0.3\textwidth}
    \includegraphics[width=\textwidth]{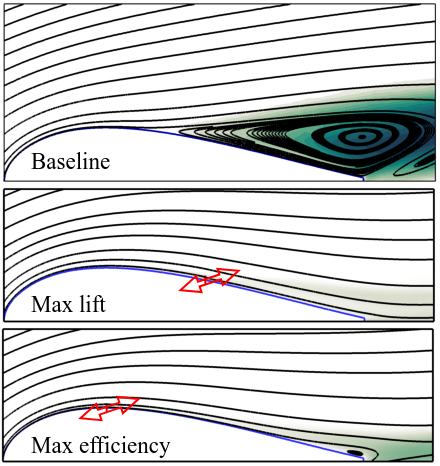}
    \caption{z/R=0.35}
    \end{subfigure}
    \caption{Streamlines of temporal average velocity field and contours of turbulence velocity at $z/R=0.25$, $0.30$ and $0.35$. The baseline case, maximum lift, and maximum efficiency cases are introduced as the top, central, and bottom row, respectively.}
    \label{fig15}
\end{figure}

To gain a deeper insight into the impact of actuation on the upper surface of the airfoil, we examine the streamlines and turbulence viscosity contours for three specific sections: $z/R = 0.25$, $0.30$, and $0.35$. These are showcased for both the baseline and actuated cases in \Cref{fig15}.
Note that, the maximum lift coefficient cases accomplish a better boundary layer reattachment than the one obtained by the maximum efficiency evaluation, although both cases drastically reduce the upper vortical structures therefore hugely modifying the forces, (increasing the lift and decreasing the drag) acting on the respective airfoils.

\begin{figure}[]
    \centering
    \begin{subfigure}[b]{0.9\textwidth}
    \includegraphics[width=\textwidth]{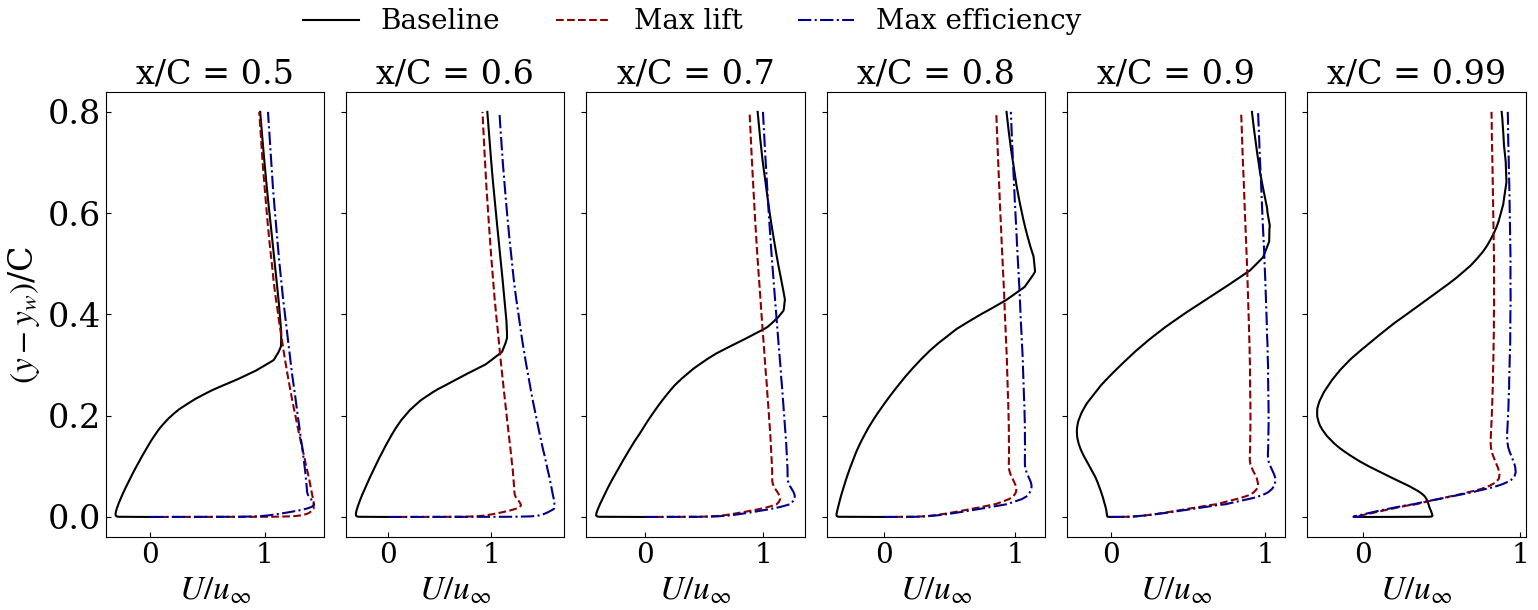}
    \caption{z/R=0.25}
    \end{subfigure}
    \begin{subfigure}[b]{0.9\textwidth}
    \includegraphics[width=\textwidth]{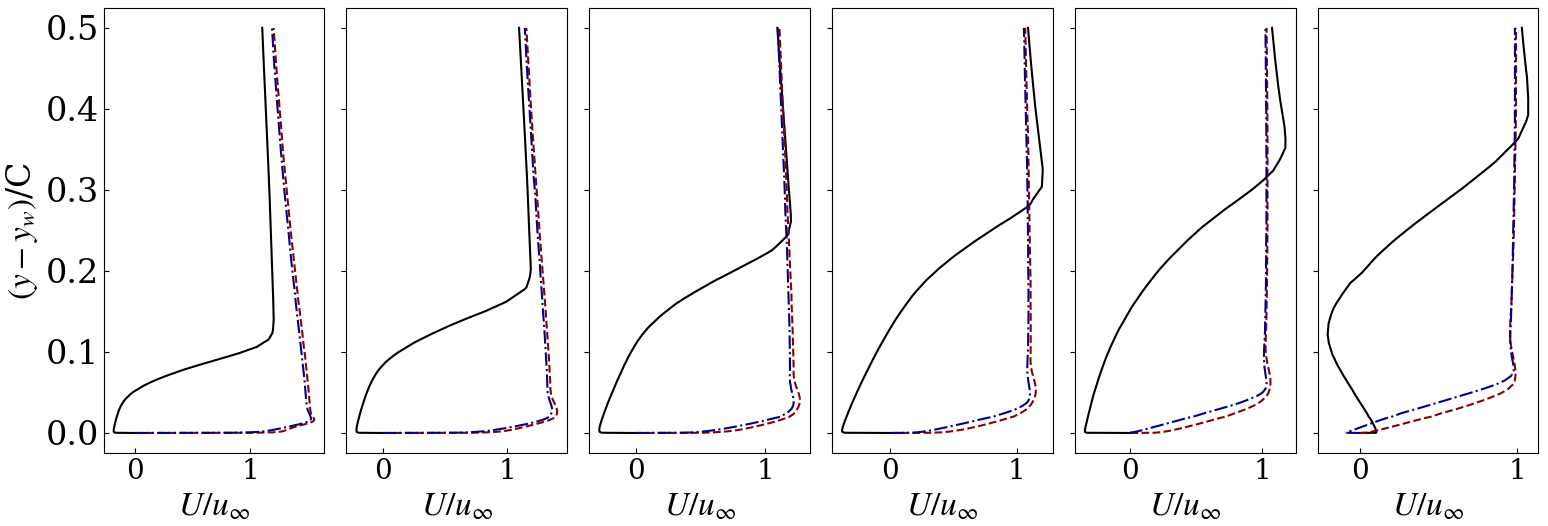}
    \caption{z/R=0.30}
    \end{subfigure}
    \begin{subfigure}[b]{0.9\textwidth}
    \includegraphics[width=\textwidth]{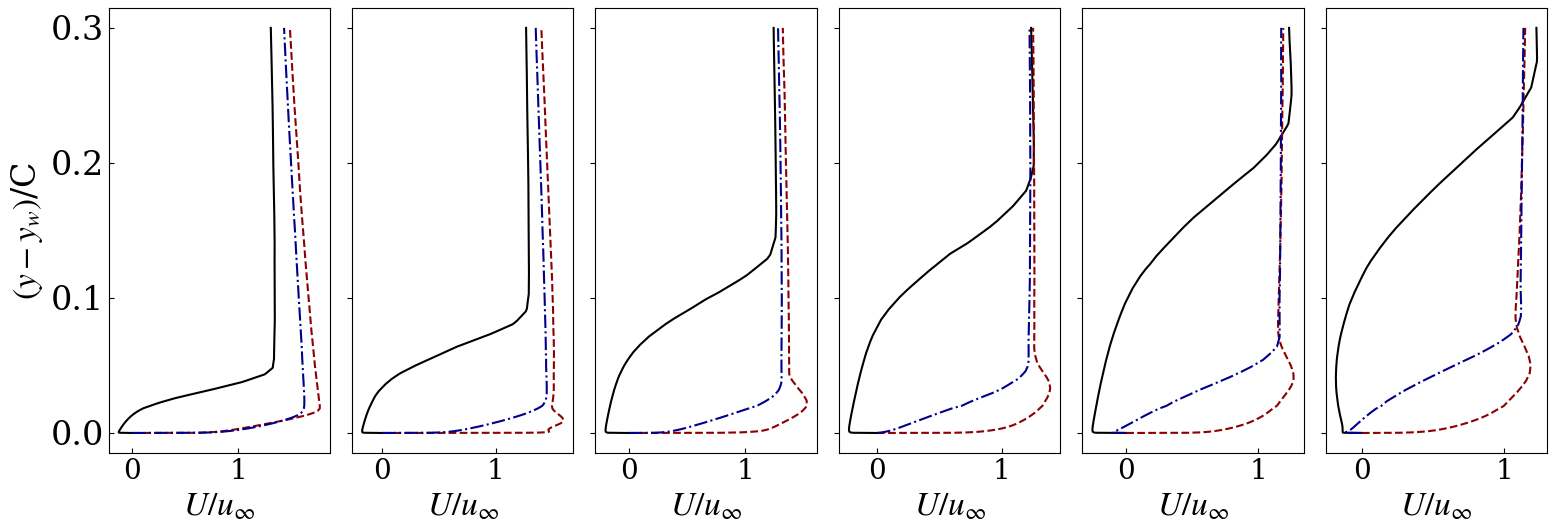}
    \caption{z/R=0.35}
    \end{subfigure}
    \caption{Normalized mean velocity magnitude profiles tangent to the airfoil surface at chord locations ranging from $x/C= 0.5$ to 1. Comparison between the baseline (black solid line), maximum lift (red dashed line) and maximum efficiency (blue dotted line) cases. (a) $z/R=0.25$; (b) $z/R=0.30$; (c) $z/R=0.35$. }
    \label{fig16}
\end{figure}

\Cref{fig16} illustrates the assessment of the normalized mean velocity magnitude profiles tangent to the airfoil surface at several chord ($x/C$) locations ranging from $x/C= 0.5$ to 1.
These profiles are depicted as a function of normalized wall-normal distance for all three airfoils evaluated. Note that for the baseline case conditions, at $x/C=0.5$ the boundary layer is already separated for all three airfoils. In accordance with what was observed in \Cref{fig15}, the boundary layer thickness is much larger at $z/R=0.25$ than at $z/R=0.35$, in fact for the highest airfoil radius, the boundary layer is just separated at this chord location. The sudden change of the normalized mean velocity magnitude profile for the baseline case at $x/C=0.99$ and for $z/R=0.25$ and $0.30$ is due to the appearance of a secondary downstream vortical structure, as clearly shown in \Cref{fig15}. For all three airfoils, once AFC is implemented, the boundary layer keeps attached along the entire chord under the maximum lift conditions, and a very small BL separation is observed at the trailing edge when maximum efficiency optimization is considered, just compare figures \Cref{fig16,fig15,fig14}. 

\begin{figure}[t]
    \centering
    \begin{subfigure}[b]{0.7\textwidth}
    \includegraphics[width=\textwidth]{optimization/s42/nut-bar.png}
    \end{subfigure}
    \begin{subfigure}[b]{0.9\textwidth}
    \includegraphics[width=\textwidth]{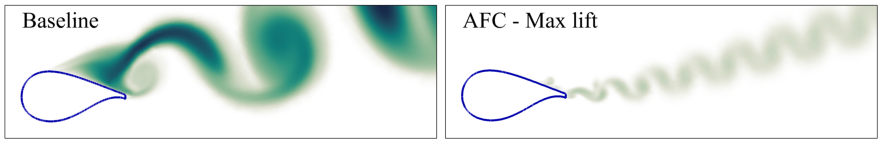}
    \caption{z/R = 0.25}
    \end{subfigure}
    \begin{subfigure}[b]{0.9\textwidth}
    \includegraphics[width=\textwidth]{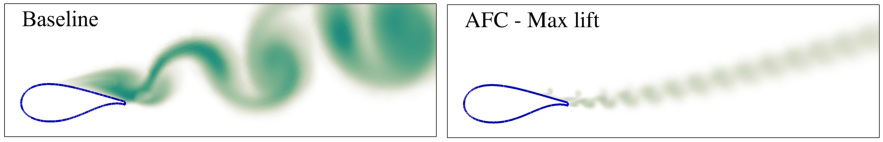}
    \caption{z/R = 0.30}
    \end{subfigure}
    \begin{subfigure}[b]{0.9\textwidth}
    \includegraphics[width=\textwidth]{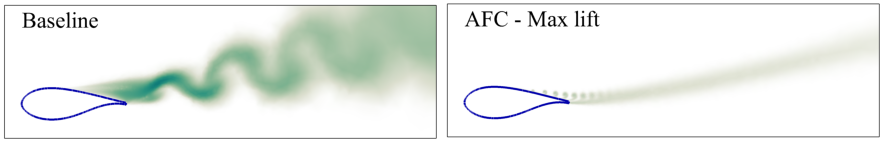}
    \caption{z/R = 0.35}
    \end{subfigure}
    \caption{Wake comparison between the baseline and the maximum lift cases for the three airfoils considered. The wake is presented as a function of the turbulence viscosity $\nu_t$ evolution.}
    \label{fig:120}
\end{figure}

In order to clarify the effect of the AFC implementation on the downstream vortical structures, \Cref{fig:120} introduces the vortex shedding generated by the baseline cases in comparison with the maximum lift ones, and as a function of the turbulence viscosity $\nu_t$. A large reduction of the dimension of the downstream structures is observed when AFC-maximum lift is considered, largely reducing as well the turbulence viscosity associated with the vortex shedding structures. This figure is in reality clarifying that, the AFC implementation is not only capable of enhancing the forces acting on the WT airfoils but it is as well minimizing the downstream wake, therefore allowing consecutive WT, in WT farms for example, to be placed more closer from each other.

\subsection{Energy assessment}

Perhaps one of the key points when willing to implement AFC in any WT resides in analyzing if the energy needed for the AFC implementation is smaller than the energy gain obtained by the turbine. This point is clarified in the present section and for each of the three airfoils optimized.

The power required to activate the synthetic jet is defined as \cite{tousi2021active,de2015comparison}:
\begin{equation}
     W_j=\frac{1}{2}\rho_jA_j\sin\theta_{\text{jet}}\overline{u^3_j}
\end{equation}

Where $\rho_j$ states for the density of the exiting fluid, $A_j$ is the cross-sectional flow area of the synthetic jets  and $\overline{u^3_j}$ is the jet time-dependent velocity profile defined as \cite{de2015comparison,de2012active}:

\begin{equation*}
    \overline{u^3_j}=\frac{1}{T/2}\int_{0}^{T/2}U_{j}^3\sin^3(2\pi ft)\mathrm{d}t=\frac{4}{3\pi}U_{j}^3
\end{equation*}

T being the SJA oscillation period.

The incremental torque represents the difference between the torque generated by the actuated section and the baseline one. It is initially essential to determine the resultant aerodynamic force ($F_{Tr}$) for both cases, which is obtained by decomposing the lift and drag forces in the plane of rotation using the inflow angle ($\varphi$). It is important to note that the provided aerodynamic coefficients are time-averaged values, ensuring a representative analysis of the aerodynamic forces.
\begin{equation*}
    F_{Tr\:AFC}=\frac{1}{2}\rho_\infty u_{rel}^2c\left ( c_{l_{AFC}}\sin\varphi- c_{d_{AFC}}\cos\varphi\right ) 
\end{equation*}
\begin{equation*}
    F_{Tr\:BSC}=\frac{1}{2}\rho_\infty u_{rel}^2c\left ( c_{l_{BCS}}\sin\varphi- c_{d_{BCS}}\cos\varphi\right ) 
\end{equation*}
By considering the forces generated in both the actuated and baseline cases, the increase in torque $\Delta Tr$ can be determined. 

\begin{equation*}
    \Delta Tr = Tr_{AFC}-Tr_{BSC} =r\cdot \left ( F_{Tr\:AFC}-F_{Tr\:BCS} \right )
\end{equation*}

At this stage, the increase in output power resulting from the lift and drag improvement can be expressed as follows, $\Omega$ denoting the angular velocity.

\begin{equation}
    \Delta W=\Delta Tr \cdot \Omega
\end{equation}
The net power balance ($W_G$) is obtained when considering the incremental power ($\Delta W$) obtained from each of the airfoils optimized and the power needed to delay the boundary layer separation  ($W_j$):
\begin{equation}
    W_G=\Delta W -W_j
    \label{eq:WG}
\end{equation}

In order to summarize the net power balance for each of the three airfoils optimized via AFC, \Cref{tab_power} is generated. The third column introduces the ratio between the jet maximum velocity $U_{j}$ versus the incoming fluid relative velocity $U_{rel}$. Notice that this ratio is kept rather constant for the three airfoils, slightly decreasing as $z/R$ increases. The last three columns represent the power per unit length required to operate the respective synthetic jets, the power increase for each airfoil due to the AFC implementation, and the net power gain generated by each airfoil under maximum lift and maximum efficiency conditions. A considerable net power gain is obtained under all conditions, the decrease of the net power gain with the increase of the $z/R$ ratio is perfectly understandable when considering that the boundary layer separates further downstream as the $z/R$ ratio increases.

\begin{table}
    \caption{Net power ratio obtained for all optimum cases}
    \centering
    \begin{tabular}{llllll}
    \hline
     z/R    &  Case & $U_{j}/U_{rel}$ & $W_J [kW]$ &  $\Delta W$ [kW] & $W_G$ [kW] \\
    \hline

    z/R=0.25    &  Max $C_l$ & 4.5 &  1.23 &  37.75 & 36.52\\
        &  Max $\eta$ & 4.6  & 1.27  & 36.29  & 35.67  \\
        \cline{2-6}

    z/R=0.30    &  Max $C_l$ & 4.5 &  1.85 &  37.91  & 36.07  \\
        &  Max $\eta$ & 4.4 & 1.77  &  36.06  &  34.3 \\
        \cline{2-6}

    z/R=0.35    &  Max $C_l$ & 4.5 &  2.63 &  25.97  &  23.35 \\
        &  Max $\eta$ & 4.3 &  2.26 &  25.45  & 23.19  \\
    \hline
    \end{tabular}
    \label{tab_power}
\end{table}

\section{Conclusion}

The paper introduces the procedure to be followed to maximize the power generated by the DTU-10MW-HAWT operating at 10 m/s. The same procedure is applicable to maximize the power generated by any HAWT at any wind speed.
The main concept behind the power increase consists in delaying the boundary layer separation or fully reattaching it to each corresponding airfoil surface.
The procedure starts when dividing the entire blade into a given number of airfoils (24 for the present application). The flow around each airfoil is studied via an independent 2D-CFD-URANS simulation. From the baseline cases the location of the boundary layer separation, its associated vortex shedding frequency, and the peak to peak amplitude are obtained. Via linking a genetic algorithms based optimizer with a mesh generator and a CFD solver, AFC-SJA optimization is then applied to each airfoil where the boundary layer is separated. From the optimization, it is obtained the five optimum AFC parameters which maximize the airfoil lift and efficiency. The results show that injection angles of 5 degrees and non dimensional groove widths of $h/C=0.005$ appear to be the most appropriate for all cases studied. Momentum coefficients around $C_\mu = 0.0098$ are as well ideal for nearly all cases studied, slightly decreasing with the $z/R$ increase. Optimum non dimensional actuation frequencies are about 4.5 times the vortex shedding one at $z/R=0.25$, but keep rising with the $z/R$ increase. The energy assessment indicates there is a considerable net power gain in each of the three airfoils optimized, therefore indicating that once the boundary layer will be reattached in all airfoils, the power generated by the WT could be largely increased.
This study proves it is possible to reattach the boundary layer on airfoils operating at very high Reynolds numbers $\mathcal{O}(10^7)$. It also highlights that if the boundary layer is reattached at all airfoils, the downstream vortical structures will sharply decrease therefore allowing to place wind turbines more close together in wind farms. The present research needs to be extended to the rest of the airfoils and to several wind speeds.

\section*{Acknowledgments}
This research was supported by the Universitat Politècnica de Catalunya under the grant OBLEA-2024, and by the Spanish Ministerio de Ciencia, Innovacion y Universidades with the project PID2023-150014OB-C21. Some of the computations were performed in the Red Española de Supercomputación (RES), Spanish supercomputer network, under the grant IM-2024-2-0008.


\appendix
\section{Baseline cases results and full Pareto front at three optimization cases}
\label{app1}

\renewcommand{\thefigure}{A\arabic{figure}}
\renewcommand{\thetable}{A\arabic{table}}
\setcounter{figure}{0}
\setcounter{table}{0}

\begin{longtable}[h]{lccccccc}
    \caption{ Main aerodynamic characteristic for the chosen airfoils of the WT blade operating at wind speed $10 m/s$\label{tab_all_sections}}\\
    \hline
     z/R    &  AoA [$^\circ$] & Airfoil type & Chord [m] & $C_l$ & $C_d$ & $x_S /C$ & $Re$ number\\
    \hline
    0.05    & $62^\circ$    & FFA-W3-600 &  5.38  & 1.26304 & 1.99744 & 0.13 & $3.4 * 10^6$ \\
    0.075   & $54.82^\circ$ & FFA-W3-600 &  5.38  & 1.70622 & 1.56704 & 0.16 & $3.9 * 10^6$ \\
    0.10    & $45.5^\circ$  & FFA-W3-600   & 5.38 & 1.82204 & 1.20326 & 0.19 & $4.7 * 10^6$ \\
    0.125   & $40.47^\circ$  & FFA-W3-600 &  5.45  & 1.65044 & 0.8074 & 0.15 & $5.6 * 10^6$ \\
    0.15    & $36.53^\circ$ & FFA-W3-600 &  5.59  & 1.75518 & 0.85115 & 0.20 & $6.6 * 10^6$ \\
    0.175   & $29.30^\circ$ & FFA-W3-480 &  5.72  & 1.59132 & 0.6584 & 0.20 & $7.5 * 10^6$ \\
    0.20    & $29.7^\circ$  & FFA-W3-480 &  5.96  & 1.58861 & 0.57023 & 0.22 & $9.1 * 10^6$ \\
    0.25    & $23.8^\circ$  & FFA-W3-480 &  6.18  & 1.58819 & 0.42856 & 0.24 & $11.0 * 10^6$ \\
    0.30    & $20.34^\circ$ & FFA-W3-360 &  6.20  & 1.57123 & 0.26995 & 0.27 & $12.7 * 10^6$ \\
    0.35    & $17.79^\circ$ & FFA-W3-301 &  6.06  & 1.56231 & 0.1162 & 0.41 & $14.1 * 10^6$ \\
    0.40    & $16.0^\circ$  & FFA-W3-301 &  5.85  & 1.40425 & 0.08035 & 0.55 & $14.8 * 10^6$ \\
    0.45    & $14.22^\circ$ & FFA-W3-301  &  5.50 & 1.30233 & 0.0486 & 0.65 & $15.3 * 10^6$ \\
    0.50    & $12.83^\circ$ & FFA-W3-301 &  5.19  & 1.3066 & 0.04569 & 0.92 & $15.3 * 10^6$ \\
    0.55    & $11.0^\circ$  & FFA-W3-301 &  4.71  & 1.20856 & 0.04464 & 1.0 & $15.2 * 10^6$ \\
    0.60    & $10.0^\circ$  & FFA-W3-301 &  4.44  & 1.20973 & 0.02964 & 1.0 & $15.2 * 10^6$ \\
    0.65    & $9.0^\circ$   & FFA-W3-241 &   3.93 & 1.21566 & 0.02985 & 1.0 & $14.8 * 10^6$ \\
    0.70    & $7.45^\circ$  & FFA-W3-241 &  3.69  & 1.22156 & 0.01684 & 1.0 & $14.3 * 10^6$ \\
    0.75    & $6.12^\circ$  & FFA-W3-241 &  3.28  & 1.24533 & 0.01568 & 1.0 & $13.7 * 10^6$ \\
    0.80    & $5.20^\circ$  & FFA-W3-241 &  2.98  & 1.25256 & 0.01459 & 1.0 & $13.1 * 10^6$ \\
    0.85    & $4.12^\circ$  & FFA-W3-241 &  2.61  & 1.20254 & 0.01379 & 1.0 & $12.4 * 10^6$ \\
    0.90    & $2.34^\circ$  & FFA-W3-241 &  2.34  & 0.94532 & 0.01266 & 1.0 & $11.6 * 10^6$ \\
    0.95    & $1.88^\circ$  & FFA-W3-241 &  2.25  & 0.71193 & 0.01197 & 1.0 & $10.8 * 10^6$ \\
    0.98    & $1.03^\circ$  & FFA-W3-241 &  1.52  & 0.50147  & 0.09261 & 1.0 & $7.5 * 10^6$ \\
    0.99    & $0.30^\circ$  & NACA - 0015 &  1.06 & 0.01839 & 0.08029 & 1.0 & $5.3 * 10^6$ \\
    \hline
\end{longtable}

\newpage
\begin{longtable}[h]{cllllllllll}
\caption{All the optimum cases of Pareto front at z/R=0.25 \label{pareto-s42-all}}\\
\hline
$F^+$ & $C_\mu$ & $\theta^\circ$ & $x_j/C$ & $h/C$ & $C_l$ & $C_l\%$ & $\eta$ & $\eta\%$ \\
\hline
4.40 & 0.0099 & 5 & 0.34 & 0.005 & 3.66 & 130.6 & 23.71 & 540 \\
4.40 & 0.0098 & 5 & 0.34 & 0.005 & 3.64 & 129.83 & 23.72 & 540.29 \\
4.40 & 0.0097 & 5 & 0.34 & 0.005 & 3.64 & 129.7 & 23.76 & 541.3 \\
4.60 & 0.0096 & 5 & 0.34 & 0.005 & 3.64 & 129.69 & 23.86 & 543.99 \\
4.60 & 0.0095 & 5 & 0.34 & 0.005 & 3.64 & 129.2 & 23.88 & 544.55 \\
4.60 & 0.0094 & 5 & 0.34 & 0.005 & 3.63 & 128.7 & 23.9 & 545.08 \\
4.60 & 0.0093 & 5 & 0.34 & 0.005 & 3.62 & 128.18 & 23.92 & 545.57 \\
4.50 & 0.0098 & 5 & 0.33 & 0.005 & 3.61 & 127.66 & 24.16 & 552.1 \\
4.60 & 0.0096 & 5 & 0.33 & 0.005 & 3.6 & 126.75 & 24.22 & 553.83 \\
4.50 & 0.0098 & 5 & 0.32 & 0.005 & 3.6 & 126.71 & 24.27 & 554.96 \\
4.70 & 0.0098 & 5 & 0.32 & 0.005 & 3.59 & 126.4 & 24.29 & 555.7 \\
4.20 & 0.0099 & 5 & 0.32 & 0.005 & 3.54 & 123.07 & 24.35 & 557.3 \\
4.40 & 0.0099 & 5 & 0.31 & 0.005 & 3.54 & 122.95 & 24.5 & 561.33 \\
4.40 & 0.0097 & 5 & 0.32 & 0.005 & 3.52 & 122.07 & 24.54 & 562.24 \\
4.50 & 0.0099 & 5 & 0.31 &0.005 & 3.49 & 120.1 & 25.08 & 576.85 \\
4.50 & 0.0098 & 5 & 0.31 & 0.005 & 3.41 & 115.09 & 25.75 & 594.88 \\
Baseline &  &   &   &   &1.58 & &3.70 &    \\
    \hline
\end{longtable}

\begin{longtable}[h]{cllllllllll}
\caption{All the optimum cases of Pareto front at z/R=0.30}\\
\hline
$F^+$ & $C_\mu$ & $\theta^\circ$ & $x/C$ & $h/C$ & $C_l$ & $C_l\%$ & $\eta$ & $\eta\%$ \\
\hline
5.41  &  0.0098  &  5  &  0.32  &  0.005  &  2.99  &  90.44  &  30.73  &  428.00 \\
5.41  &  0.0097  &  5  &  0.32  &  0.005  &  2.98  &  89.80  &  30.78  &  428.86 \\
5.64  &  0.0099  &  5  &  0.30  &  0.005  &  2.97  &  89.17  &  31.03  &  433.16 \\
5.64  &  0.0098  &  5  &  0.31  &  0.005  &  2.97  &  89.17  &  31.08  &  434.02 \\
5.64  &  0.0096  &  5  &  0.31  &  0.005  &  2.95  &  87.89  &  31.16  &  435.39 \\
3.1  &  0.0099  &  5  &  0.31  &  0.005  &  2.95  &  87.89  &  31.24  &  436.76 \\
3.1  &  0.0097  &  5  &  0.31  &  0.005  &  2.94  &  87.26  &  31.31  &  437.97 \\
5.41  &  0.0096  &  5  &  0.31  &  0.005  &  2.94  &  87.26  &  31.63  &  443.47 \\
5.41  &  0.0095  &  5  &  0.31  &  0.005  &  2.93  &  86.62  &  31.67  &  444.15 \\
3.12  &  0.0095  &  5  &  0.30  &  0.005  &  2.91  &  85.35  &  31.78  &  446.04 \\
3.01  &  0.0095  &  5  &  0.30  &  0.005  &  2.91  &  85.35  &  31.81  &  446.56 \\
3.38  &  0.0095  &  5  &  0.30  &  0.005  &  2.91  &  85.35  &  31.85  &  447.29 \\
3.41  &  0.0095  &  5  &  0.30  &  0.005  &  2.91  &  85.35  &  31.9  &  448.10 \\
3.12  &  0.0095  &  5  &  0.30  &  0.005  &  2.91  &  85.35  &  31.92  &  448.45 \\
3.07  &  0.0098  &  5  &  0.29  &  0.005  &  2.89  &  84.07  &  32.05  &  450.68 \\
3.07  &  0.0097  &  5  &  0.29  &  0.005  &  2.89  &  84.07  &  32.06  &  450.85 \\
3.12  &  0.0096  &  5  &  0.29  &  0.005  &  2.88  &  83.43  &  32.08  &  451.20 \\
3.12  &  0.0095  &  5  &  0.29  &  0.005  &  2.87  &  82.80  &  32.09  &  451.37 \\
3.12  &  0.0098  &  5  &  0.29  &  0.005  &  2.85  &  81.52  &  32.25  &  454.12 \\
5.64  &  0.0093  &  5  &  0.29  &  0.005  &  2.84  &  80.89  &  32.47  &  457.90 \\
Baseline &  &   &   &   &1.57 & &5.82 &    \\
    \hline
    \label{pareto-s48-all}
\end{longtable}

\newpage
\begin{longtable}[h]{cllllllllll}
    \caption{All the optimum cases of Pareto front at z/R=0.35}\\
    \hline
     $F^+$ & $C_\mu$ & $\theta^\circ$ & $x/C$ & $h/C$ & $C_l$ & $C_l\%$ & $\eta$ & $\eta\%$ \\
9.51  &  0.0093  &  5  &  0.58  &  0.005  &  2.63  &  68.94  &  28.43  &  100.52 \\
4.97  &  0.0093  &  5  &  0.58  &  0.005  &  2.63  &  68.85  &  28.44  &  100.61 \\
9.43  &  0.0092  &  5  &  0.57  &  0.005  &  2.63  &  68.69  &  28.55  &  101.34 \\
9.63  &  0.0092  &  5  &  0.56  &  0.005  &  2.63  &  68.59  &  28.74 &  102.70 \\
9.63  &  0.0093  &  5  &  0.54  &  0.005  &  2.62  &  68.47  &  29.08  &  105.11 \\
9.43  &  0.0092  &  5  &  0.54  &  0.005  &  2.62  &  68.37  &  29.16  &  105.69 \\
4.86  &  0.0084  &  5  &  0.57  &  0.005  &  2.61  &  67.81  &  29.30  &  106.67 \\
9.94 &  0.0083  &  5  &  0.58  &  0.005  &  2.61  &  67.56  &  29.31  &  106.71 \\
9.63  &  0.0082  &  5  &  0.58  &  0.005  &  2.61  &  67.54  &  29.39  &  107.31 \\
9.23  &  0.0082  &  5  &  0.58  &  0.005  &  2.61  &  67.51  &  29.40  &  107.40 \\
9.63  &  0.0082  &  5  &  0.57  &  0.005  &  2.61  &  67.50  &  29.42  &  107.52 \\
9.23 &  0.0081  &  5  &  0.58  &  0.005  &  2.61  &  67.33  &  29.49  &  107.97 \\
9.02  &  0.008  &  5  &  0.58  &  0.005  &  2.60  &  67.28  &  29.53  &  108.26 \\
9.02  &  0.0079  &  5  &  0.58  &  0.005  &  2.60  &  67.27  &  29.66  &  109.21 \\
8.92  &  0.0079  &  5  &  0.58  &  0.005  &  2.60  &  67.09  &  29.67 &  109.29 \\
8.92  &  0.0078  &  5  &  0.58  &  0.005  &  2.60  &  67.01  &  29.73  &  109.71 \\
5.07 &  0.0075  &  5  &  0.58  &  0.005  &  2.59  &  66.61  &  30.03  &  111.83 \\
9.73 &  0.0074  &  5  &  0.58  &  0.005  &  2.59  &  66.49  &  30.13  &  112.49 \\
8.31 &  0.0073  &  5  &  0.58  &  0.005  &  2.59  &  66.37  &  30.20  &  113.03 \\
9.63 &  0.0071  &  5  &  0.58  &  0.005  &  2.59  &  66.30  &  30.37  &  114.22 \\
9.73 &  0.0081  &  6  &  0.58  &  0.005  &  2.58  &  65.82  &  30.42  &  114.54 \\
9.73 &  0.0094  &  5  &  0.37  &  0.005  &  2.58  &  65.69  &  33.45  &  135.89 \\
9.73 &  0.0095  &  5  &  0.36  &  0.005  &  2.58  &  65.50  &  33.78  &  138.22 \\
9.02 &  0.0089  &  5  &  0.36  &  0.005  &  2.57  &  64.80  &  34.07  &  140.31 \\
6.23  &  0.0082  &  5  &  0.37  &  0.005  &  2.56  &  64.17  &  34.11  &  140.56 \\
9.63  &  0.0082  &  5  &  0.33  &  0.005  &  2.51  &  61.15  &  35.07  &  147.38 \\
9.73  &  0.0074  &  5  &  0.33  &  0.005  &  2.49  &  60.23  &  35.90  &  153.18 \\
Baseline &  &   &   &   &1.56 & &14.18 &    \\
    \hline
    \label{pareto-s54-all}
\end{longtable}


\newpage

  \bibliographystyle{elsarticle-num} 
  \bibliography{bib}



\end{document}